\documentclass[traditabstract,usenatbib]{aa}
\usepackage{amsmath}
\usepackage{bm}
\usepackage{txfonts}
\usepackage{graphicx}                      
\usepackage{subfigure}                      
\usepackage{url} 
\usepackage{natbib}               
\usepackage{color}
\bibpunct{(}{)}{;}{a}{}{,}	

\newcommand{\nnn}[1]{{\color{black} #1}}

\begin{document}
   \title{ Isolated massive stars in the Galactic center: The dynamic contribution from the Arches and Quintuplet star clusters.}


\author{
M. ~Habibi \inst{\ref{inst1},\ref{inst3} }, Andrea Stolte  \inst{\ref{inst1} }, Stefan Harfst \inst{\ref{inst2} }
}
\authorrunning{ Habibi et al.}

\institute{
	   Argelander Institut f\"ur Astronomie, Universit\"at Bonn, Auf dem H\"ugel 71, 53121 Bonn, Germany \email{[mhabibi;astolte]@astro.uni-bonn.de}\label{inst1}
	   \and  Technische Universit\"at Berlin, Zentrum f{\"u}r Astronomie und Astrophysik, Hardenbergstra\ss e 36, 10623 Berlin, Germany\email{harfst@astro.physik.tu-berlin.de} \label{inst2}
	   \and  Member of the International Max Planck Research School (IMPRS) for Astronomy and Astrophysics at the Universities of Bonn and Cologne. \label{inst3}
	   	   }

  \date{Received..; accepted ...}


  \abstract
   {Recently, more than 100 Wolf-Rayet and OB stars were identified in the Galactic center. \nnn{About a third of these sources} are not spatially associated with any of the known star clusters in this region. We probe the distribution of drifted sources in numerical models of the massive clusters in the Galactic center and compare it to the observed distribution of isolated massive sources in this region. We find that stars as massive as $100\,M_\odot$ drift away from the center of each cluster by up to $\sim\,60$ pc using the cluster models. \nnn{Our best model reproduces $\sim60\%$ of the known isolated massive stars  out to $80 \,\mathrm{pc}$ from the center of the Arches cluster. This number increases to $70-80\%$ when we only consider the region of $\sim 20$ pc from the Arches cluster. }}

   \keywords{Galaxy: center, open clusters and associations--: individual: 
   Arches, Quintuplet, -- stars: early-type--stars: Wolf-Rayet--stars: kinematics and dynamics -- methods: N-body simulations }

\titlerunning{Isolated massive population in the Galactic center.}
\authorrunning{M. Habibi et al.}

   \maketitle
%

\section{Introduction}

Galactic nuclei are ideal laboratories to investigate star formation in extreme conditions such as a strong tidal field, high UV radiation and a strong magnetic field.
The only galactic nucleus where we can resolve the stellar population into individual stars is the center of our Galaxy at a distance of $\sim8.0\,\mathrm{kpc}$ (\citealt{ghez2008}, \citealt{gillessen}). However, the conditions for star formation and the dynamics of this region yet need to be understood as it harbors dense molecular clouds, a high star formation rate per unit volume, and the largest concentration of massive stars and star clusters in the Milky Way (e. g. \citealt{1996ARA&A..34..645M}; \citealt{2002ASPC..285..381F};  \citealt{2007A&A...467..611F}).

The GC region hosts three  starburst clusters with masses in excess of $\sim 10^4\,M_{\odot}$ and core radii of $\sim 0.15-1\,\mathrm{pc}$ (\citealt{1993ApJ...407L..77E}; \citealt{figer2002}; \citealt{Espinoza}). These three compact and massive clusters are the Young Nuclear Cluster surrounding the supermassive black hole, as well as the Quintuplet and the Arches clusters. Recent observations of isolated sources in the GC region revealed that, similar to these three clusters, the field stars in this area encompass many massive sources (\citealt{2011MNRAS.417..114D}; \citealt{Mauerhan2010_main}; \citealt{Mauerhan_matching}; \citealt{Wang2010}). A population of distributed very massive Wolf-Rayet stars with initial masses in excess of $20-40\,M_{\odot}$ were detected  within a few dozen of pc from the super-massive black hole, Sgr~A*, by X-ray observations, accompanied by spectroscopic studies and Paschen-$\alpha$ (Pa$\alpha$) narrow-band imaging with HST (\citealt{Wang2010}; \citealt{2011MNRAS.417..114D}).  Up to now, more than 100 Wolf-Rayet stars and O supergiants have been spectroscopically identified in the Galactic center region \citep{Mauerhan2010_main}, including the known cluster members.

 As about a third of these sources are located outside the three massive starburst clusters, they were suggested to provide evidence for {\sl isolated} high-mass star formation in the GC (\citealt{2011ASPC..439..104D}, \citealt{2013arXiv1309.7651O}). Observations of massive stars in the solar neighborhood show that generally massive stars form in groups and associations (\citealt{2003ARA&A..41...57L};  \citealt{2007ARAA..45..481Z}; \citealt{2008AA...490.1071G}). But it is not clear if we can generalize these findings to different galactic environments. 
On the other hand, dynamical evolution of stellar populations in the  GC region can become dramatic under the strong effect of the GC tidal field. If so, dense and massive clusters like the Arches and Quintuplet can shape the distribution of the field stars in the region.

 In fact, the Arches and Quintuplet clusters are observed to be already mass segregated at ages of $2-6\,\mathrm{Myr}$.
A recent study by \cite{benjamin} showed that the Quintuplet cluster at an age of 3-5$\,\mathrm{Myr}$ exhibits a flat mass function slope of  $ -1.68 \pm 0.1$ in the cluster center compared to the standard \cite{Salpeter} initial mass function (IMF) of $-2.3$. A similarly flat mass function was found in the central region of the
 Arches cluster,  but the slope increases substantially  towards larger radii (\nnn{\citealt{1999ApJ...525..750F}}, \citealt{stolte_2005}, \citealt{Espinoza}, \citealt{habibi}).
 A study by  \cite{Harfst2010} implemented N-body simulations of the Arches cluster to investigate the internal dynamical evolution of the cluster. By comparing their models to the observational data from the central 0.4 pc of the Arches cluster, they could constrain the initial conditions and construct a dynamical model of the Arches cluster that best represented the central stellar mass distribution, hereafter the best-fitting model of the Arches cluster. From this model, a steep increase of the stellar mass function slope as a function of the cluster center distance was predicted. 
The Arches cluster was later studied by \cite{habibi} to a larger radius of $\sim 1.5\,\mathrm{pc}$. Observations of the high-mass part of
the mass spectrum in this study ($M >\sim 10M_{\odot}$) revealed a depletion of massive stars in the cluster outskirts. 
 In this previous study, we compared the measured slope of the mass function to the slopes predicted by the N-body simulations in different annuli out to the tidal radius of the Arches cluster. 
  This comparison showed that the Arches cluster exhibits characteristics of a normal, i. e. with a normal initial mass function, but dynamically evolved cluster. The dynamical evolution of the Arches and Quintuplet clusters, however, not only changes the distribution of stars inside the clusters, it can also change the distribution of field stars in the GC region. Through gravitational interactions between stars in dense and compact clusters stars can accelerate to become runaways (e.g. \citealt{1967BOTT....4...86P}; \citealt{1986ApJS...61..419G}). Moreover, the dynamical evolution of clusters under the influence of the Galactic tidal field leads to the formation of  tidal arms. These tidal structures are mostly observed for globular clusters which evolve for many Gyrs (e.g. \citealt{2001ApJ...548L.165O}).  As massive clusters in the GC dissolve within a few Myrs (\citealt{2000ApJ...545..301K}; \citealt{2001ApJ...546L.101P}), dynamical evolution under the effects of the strong tidal field of the GC leads to the formation of extended tidal structures during shorter timescales. These tidal structures, in turn, can significantly contribute to the field stars in the GC region.

In this study, we analyze the best-fitting Arches model presented by \cite{Harfst2010}, extended to incorporate the effect of the Galactic
center tidal field, to investigate the contribution of the Arches and Quintuplet clusters to the observed population of isolated massive stars detected by \citep{Mauerhan2010_main}. This paper is organized as follows: In Sect. \ref{sec: data}, a summary of observational studies to detect massive stars in the GC region, together with our criteria to construct an observational reference sample, is presented. In Sect. \ref{sec: simul}, we describe the computational methods, a grid of different models based on distinct physical assumptions, and a method to find the best-matching model to reproduce the observations. In Sect. \ref{sec: results}, we analyze this best-matching model to derive the spatial distribution of the massive drifted sources compared to the observed population. The velocity distribution of the drifted sources from the modeled clusters and their expected spatial and mass distributions are also predicted in this section. A summary of our findings is presented in Sect.\ref{sec:conc}.

\section{The observed population of Wolf-Rayet Stars and O Supergiants}
\label{sec: data}
A recent  HST/NICMOS  Pa$\alpha$ survey of the Galactic center was carried out by \cite{Wang2010} in order to study the distribution of young massive stars in the GC region. The survey covered the central region of the Galaxy, $\approx0.65^\circ\times0.25^\circ (l,b)$. At the distance of the GC from the Sun ($8\,\mathrm{Kpc}$; \citealt{ghez2008}, \citealt{gillessen}) an angular distance of $1^{\circ}$   corresponds to a projected distance of $\approx 140\,\mathrm{pc}$, such that the survey covers an area of $91 \times 35\, \mathrm{pc^2}$. \cite{Wang2010} employed two narrow band filters, F187 (1.87 $\mu m$) and F190 (1.90 $\mu m$), to detect sources with  Pa$\alpha$ excess. The Pa$\alpha$  emission line (1.87 $\mu m$) is the strongest line in the infrared band, which mainly originates from warm and hot ionized gas. Accordingly, Pa$\alpha$ point sources are chiefly produced by evolved massive stars \citep{Mauerhan2010_main}. Later, \cite{Mauerhan2010_main} performed a follow-up near-infrared spectroscopic survey targeting the strongest Pa$\alpha$ sources with bright $K_{s}$-band counterparts ($K_{s}< 12.7$ mag). For this reason  \cite{Mauerhan2010_main} employed the  $\mathrm{JHK_{s}}$ photometry down to 15.6 mag, taken from the 2MASS and SIRIUS catalogs (2MASS: Two Micron All Sky Survey; \citealt{2mass}, SIRIUS: Simultaneous three-color InfraRed Imager for Unbiased Surveys; \citealt{sirius}). These magnitude ranges cover the predicted magnitude for WR stars in the GC region, which is dominated by WNL stars ($K<12$; \citealt{figer_thesis}). Fainter WNE stars are still detectable due to their strong Pa$\alpha$ emission. Therefore, their sample of WN stars in the Pa$\alpha$ survey area  is nearly complete whereas the provided WC sample is presumably not as complete (see Sec 6.1 of \cite{Mauerhan2010_main} for a detailed discussion).

In this study, we use the catalog of isolated massive sources in the GC provided by \cite{Mauerhan2010_main} to perform a statistical analysis.
\cite{Mauerhan2010_main} present a catalog of all the identified evolved massive stars in the GC including their  newly discovered sources, those previously identified in the Arches, Quintuplet and the Nuclear star clusters as well as those found outside the Pa$\alpha$ survey coverage.  In the \nnn{presented} catalog \nnn{by \cite{Mauerhan2010_main}}, sources that lie outside  the Pa$\alpha$ survey area (located at galactic longitudes of $0.2^{\circ}\lesssim l \lesssim0.6^{\circ}$) are taken from the study by \cite{Mauerhan_matching}.

 \nnn{\cite{Mauerhan_matching} performed} a spectroscopic study on a pre-compiled list of  potential near-infrared counterparts to hard X-ray sources. They derived the candidate list by matching the Chandra catalog of X-ray sources in the central $2^\circ \times 0.8^\circ (l,b)$ of the galaxy \citep{chandra} to the available near-infrared surveys in the region. However, \nnn{the studied sample by \cite{Mauerhan_matching} in the region outside the Pa$\alpha$ survey area} suffers from selection bias. For the spectroscopic follow-up \cite{Mauerhan_matching} give precedence to the sources close to previously observed diffuse mid-infrared structures.  Diffuse infrared emission is the characteristic of young star forming regions in the GC. \nnn{\cite{Mauerhan_matching} mostly studied such regions  which may harbor ongoing star formation with the hope} to enhance the detection rate of massive stars. To avoid the selection bias, we limit our sample to the region that is covered by the Pa$\alpha$ survey by selecting sources at longitudes \nnn{$l< 0.23^\circ$ and latitudes $b>-0.1^\circ$} from their catalog (see Fig. \ref{dens_map}). We utilize this truncated list of observed isolated massive sources as our observational reference sample\footnote{ \nnn{A more recent study by \cite{2012MNRAS.425..884D} includes fainter massive field stars in the GC ($Ks \gtrsim 13$ mag). Their catalog is not employed in our study as most of these  stars still do not have available spectroscopic identifications. Moreover, as they have not performed a spectroscopic survey, the spatial distribution of sources with previous spectroscopic identifications is probably biased to the target area of earlier studies.}}. The final employed list of 35 observed isolated massive stars in the GC region is presented with positions and spectral types
in Table \ref{source_list}, along with their reference studies. \nnn{This table contains 13 WN stars, 11 WC stars, 1 LBV star, and 10 OB supergiants outside the three clusters in the Pa$\alpha$ survey region (see Fig. \ref{dens_map}).}

\section{ Dynamical cluster model}
\label{sec: simul}
\subsection{N-body simulations}

The dynamical cluster models presented here are based on the work by \citet{Harfst2010} who compared in detail the results of numerical simulations with the observational data from the central $0.4\,\mathrm{pc}$ of the Arches cluster \citep{stolte_2005}. They  constrained the initial conditions and found a best-fitting Arches model. This best-fitting Arches model is used here as a starting point for the simulations. 

The setup of the dynamical models of the Arches cluster as described in the following is similar to the models presented by \cite{2012ApJ...756..123O}.
For our simulations, the age of the Arches cluster is set to $t_\mathrm{age} = 2.5\mathrm{\,Myr}$ (\citealt{blum}; \citealt{Najarro}; \citealt{martins}), and we assume that the cluster is initially (at $t=0$\,Myr) gas-free and in virial equilibrium and that its mass is distributed following a King profile \citep{1966AJ.....71...64K}. A single-aged
 stellar population is used and no initial mass segregation is taken into account.

According to the best-fit model of \citet{Harfst2010}, the observed present-day mass function of the Arches cluster is the result of a dynamically evolved Salpeter IMF  (\citealt{Salpeter}) at the given age of the cluster. The total mass of the cluster is then given by the total number of massive stars (stars with $m> 10\,M_{\odot}$ which are complete in the observational data) and the lower mass limit of the IMF. For the latter, we adopt $m_\mathrm{low} = 0.5\,M_{\odot}$. As a result, the total mass of the cluster model is $M=4.8\cdot10^4\,M_{\odot}$ which corresponds to an initial number of massive stars  of $N_{m>10 M_{\odot}} \approx 500$.

The King profile concentration parameter was set to $W_0 = 3$. The initial size of the cluster is given by the initial virial radius, which was set to $0.77\,\mathrm{pc}$ (see \citealt{Harfst2010}). The corresponding core radius is about $0.4\,\mathrm{pc}$ initially, and shrinks, due the dynamical evolution of the cluster, down to the observed $0.2\,\mathrm{pc}$ \citep{Espinoza}. 

Stellar evolution and the orbit of the Arches cluster in the Galactic center potential have been neglected in the best-fit model of \citet{Harfst2010}. Here we extend that model, as described below, to include these two effects. The simulations were carried out with the direct $N$-body code {\tt kira} from the $starlab$-package (\citealt{1996ASPC...90..413M}; \citealt{2001MNRAS.321..199P}; \citealt{2003IAUS..208..331H}), which includes modules for stellar evolution and an external potential. The stellar evolution module includes mass loss by stellar winds and binary evolution (\citealt{1989ApJ...347..998E}; \citealt{1996A&A...309..179P}; \citealt{1997MNRAS.291..732T}; \citealt{1998A&A...329..551L}). The external potential that was used in our simulations is that of a power-law mass distribution with $M_\mathrm{gal}(r) = 4.25\times 10^6 (r/\mathrm{pc})^{1.2}\,M_{\odot}$ and is based on \citet{1999A&A...348..457M}.

The cluster orbit in the potential is given by the six phase space coordinates composed of the 3D position and velocity  of the cluster at any given time. For the present day ($t=2.5$ Myr), we know the line-of-sight velocity ($95\,\mathrm{km\,s^{-1}}$, \citealt{figer2002}) and the projected position ($26\,\mathrm{pc}$ from the Galactic center). Additionally, \citet{2008ApJ...675.1278S} determined the proper motion of the Arches cluster and found a 2D velocity of $212\,\mathrm{km\,s^{-1}}$ (anti-)parallel to the Galactic plane. The proper motion of the cluster was later revised to a slightly lower value of $172 \pm 15\, \mathrm{km s^{-1}}$ (see \cite{2012ApJ...751..132C} for details). We have therefore adopted the mean present-day 2D velocity 
     of 190 $\mathrm{km s^{-1}}$ for our cluster simulations. As the difference in the 3D space motion is small, we do not expect the orbit to change substantially. If we define a coordinate system in which the $x$-axis is along the Galactic plane, the $y$-axis along the line-of-sight, and the $z$ axis towards the Galactic north pole, we obtain the following vectors for the position, $\boldsymbol{r}_\mathrm{cluster}$, and velocity, $\boldsymbol{v}_\mathrm{cluster}$,  of the cluster at the present epoch:
\begin{equation}
\begin{aligned}
\boldsymbol{r}_\mathrm{cluster} = (-24,\; R_\mathrm{GC} + d_\mathrm{los}, 10)\, \mathrm{pc} \\
\\
\boldsymbol{v}_\mathrm{cluster} = (-190, 95, 0)\, \mathrm{km}\,\mathrm{s}^{-1}
\end{aligned}
\end{equation}
where the not well-known $d_\mathrm{los}$ determines the distance of the cluster to the Galactic center along the line of sight. \cite{andrea2008} discussed the formation of the Arches cluster inside the central $200$ pc of the Galaxy as one of the most likely formation scenarios. Furthermore, considering its orbital properties and the measured foreground extinction of the cluster, which is lower than the extinction found towards the central parsec, \cite{andrea2008}  concluded that the cluster is likely in front of the GC today. For the simulations, we therefore assume a line-of-sight location of $d_\mathrm{los} = -100\mathrm{\,pc}$ in front of the GC, at its present age of 2.5 Myr.

The present-day position and velocity can be used to numerically integerate the orbit backwards in time to find the initial position and velocity at $t=0$ Myr. From there, the full cluster was  integrated in {\tt kira} until $t=6$\,Myr  including the effects of stellar evolution and the GC tidal field. A total of ten random realizations of this model were integrated in order to allow for statistical analysis. A full snapshot of the cluster (mass, position, and velocity for every star) was stored every $\sim0.5$\,Myr. From these snapshots the projection on the plane of the sky is easily obtained.

\subsection{Model grid }\label{sec:models_grid}

 \nnn{ The Arches and Quintuplet clusters share similar characteristics. Other than their proximity, observations yield similar estimates of the present-day mass of the clusters (\cite{2014PhDT.........1H}, \cite{habibi}, \cite{2012ApJ...751..132C}). Additionally,   preliminary results on the proper motion study of the Quintuplet cluster suggests that the motion of the Quintuplet cluster is similar to the orbital motion of the Arches cluster \citep{2011sca..conf..304S}. However, the Quintuplet is slightly older than the Arches (2-3 Myr vs 3-5 Myr, \cite{Najarro} and \cite{Figer1999}, respectively) and appears to be more dispersed than the Arches cluster. Quintuplet is more than 100 times less dense than  the Arches cluster \citep{Figer2008}. These properties are suggestive to consider the Quintuplet cluster as an older representation of the Arches cluster.

In order to model the observed population of massive field stars in the GC region, we need to consider the contribution of both the Arches and the Quintuplet clusters.
Our simulated cluster provides the contribution from the Arches cluster. We assume the Quintuplet cluster is a snapshot of the Arches cluster at older ages as it evolves in the GC region.} In order to construct a single model that contains both the Arches and Quintuplet clusters, the data from the two simulated clusters are combined such that their projected distance on the plane of the sky matches the observations. Assuming the GC distance of $8\,\mathrm{kpc}$, the observed separation of the clusters along the Galactic plane and the Galactic north pole is $\sim 6\,\mathrm{pc}$ and $\sim 11\,\mathrm{pc}$, respectively.

Our models \nnn{do not include} the Nuclear cluster since we do not expect that ejected sources from the Nuclear cluster linger in the GC region. \cite{schodel} observed the proper motions of stars within a distance of $1.0\,\mathrm{pc}$ from Sgr~A*. They found  few stars with proper motion velocities which exceed 400 $\mathrm{km\,s}^{-1}$ in both  the radial and tangential axes. Out of these candidates only one source has a velocity uncertainty below $1\sigma$. It is noteworthy that close to the center of the Nuclear cluster the black hole governs the dynamics. However, as first pointed out by \cite{hills} mechanisms like  interaction of a massive binary with the black hole produce ejectors with hyper-velocities of $\gtrsim 1000\, \mathrm{km\,s}^{-1}$. These hyper-velocity ejecta traverse our target region of $50\,\mathrm{pc}$ around the GC in $\sim 10^{4}$ yrs. Comparing this value to the maximum predicted ejection rate of $10^{-4} \mathrm{yr}^{-1}$ (\citet{2003ApJ...599.1129Y}; \citet{2007ApJ...656..709P}) we expect $\sim 1$ hyper-velocity star originated from the Nuclear cluster to contribute to the observed sample of massive stars in the GC at any given time. 

To obtain an estimate of the number of drifted sources\footnote{\nnn{We consider sources  outside the tidal radius ($\sim 1.6\,\mathrm{pc}$; \citealt{habibi}) of the Arches and Quintuplet clusters (\citealt{2014PhDT.........1H}) that are part of the tidal arms as drifted sources.}} which are expected to originate from the Quintuplet, we evolve the model for a longer time. The age of the Arches cluster is estimated to be 2.5 Myr. The age of the Quintuplet cluster, however, is less constrained. \cite{Figer1999} conducted a photometric and spectroscopic study on  massive stars in the cluster and derived an age of 3 to 5 Myr based on the types of the stars they found. A more recent spectral analysis of WN stars by  \cite{LiermannWN} favors a younger age of 3 million years, while on the HR diagram OB stars in the cluster populate a 4 Myr isochrone (\citealt{LiermannHR}). To account for the age uncertainty of the Quintuplet cluster, we use three different snapshots of our model at 4, 4.5 and 5 Myr to represent the current extended population originating from the Quintuplet. Each of the three snapshots are combined with the Arches model at 2.5 Myr to estimate the total number of isolated massive stars originating from both clusters in the GC today.

Another parameter which has a significant effect on the predicted number of WR stars in the region is the minimum initial mass of a WR progenitor star. A WR star represents a massive star at its late evolutionary stage. \cite{2004MNRAS.353...87E} concluded that at solar metallicity the minimum initial mass of a WR progenitor is $\sim 25 \,M_{\odot}$. However, modeling WR stars is still a topic of active research. Recent Geneva stellar evolution tracks \citep{2012A&A...537A.146E} determine a minimum initial mass of $32\, M_{\odot}$ for a He-burning star with an age of 5.26 Myr. In these models the minimum initial mass of a 4.97 Myr star, which is at its core He-exhaustion phase, is $40 \,M_{\odot}$. For rotating models these numbers are higher and are close to $60\, M_{\odot}$. Since, since observed WRs are defined by their prominent broad emission lines, the spectral analysis of WR stars is only possible through detailed model atmospheres. \cite{2006A&A...457.1015H} performed a spectral analysis on a large sample of Galactic WN stars using the Potsdam Wolf-Rayet (PoWR) model atmosphere code. Based on the rough qualitative agreement of their analysis and available non rotating Geneva tracks at the time \citep{Geneva2003} they found a minimum initial mass of $37\,M_{\odot}$ for a star to reach any WR phase. Later  \cite{2012ASPC..465..243S}  studied Galactic WC stars using  the same model atmosphere code. They conclude that these stars are evolved from progenitors of 20 to $45\,M_{\odot}$. On the basis of current knowledge of evolved massive stars we consider 3 different minimum initial masses of 20, 40, and $60\, M_{\odot}$ for a star to eventually evolve into a WR star.

Based on these different assumptions about the age of the Quintuplet cluster and the initial mass of the WR progenitors, we construct nine models of massive stars in the Arches and Quintuplet clusters. We compare the population of drifted massive stars of the nine models with the observed isolated massive sources (see Table \ref{h_d}).

\subsection{Comparison of the observed distribution of isolated high-mass stars with the model grids}
\subsubsection{The concept of the histogram difference}

Constructing a grid of different models (see Sect. \ref{sec:models_grid}), we are interested in finding the most similar model to the observed population. In order to measure the deviation of the models from the observed data, histograms of the data and each model are constructed by calculating spatial distances of all the stars, including the Quintuplet members, to the center of the Arches cluster as a reference point. One way to quantify histogram differences is to use the number count difference in each bin which is equivalent to calculating the classic Euclidean distance between the feature vectors\footnote{ A feature vector is a vector for which the components of the vector represent properties of some data set. Here, constructed histograms represent the feature vectors of each distribution.} of two distributions. The square of the classic Euclidean distance of two N-dimensional vectors $\mathbf{P}$ and $\mathbf{Q}$  is defined as follows:

$$ D_{euclid}^{2}(\mathbf{P},\mathbf{Q})=(\mathbf{P}-\mathbf{Q})\cdot(\mathbf{P}-\mathbf{Q})^T=\sum_{i=1}^{N} (p_{i} -q_{i})^2$$

\begin{figure*}[t] 
\centering $
\begin{array}{ccc}
   \includegraphics[trim=10mm 5mm 5mm 8mm, clip,scale=0.5]{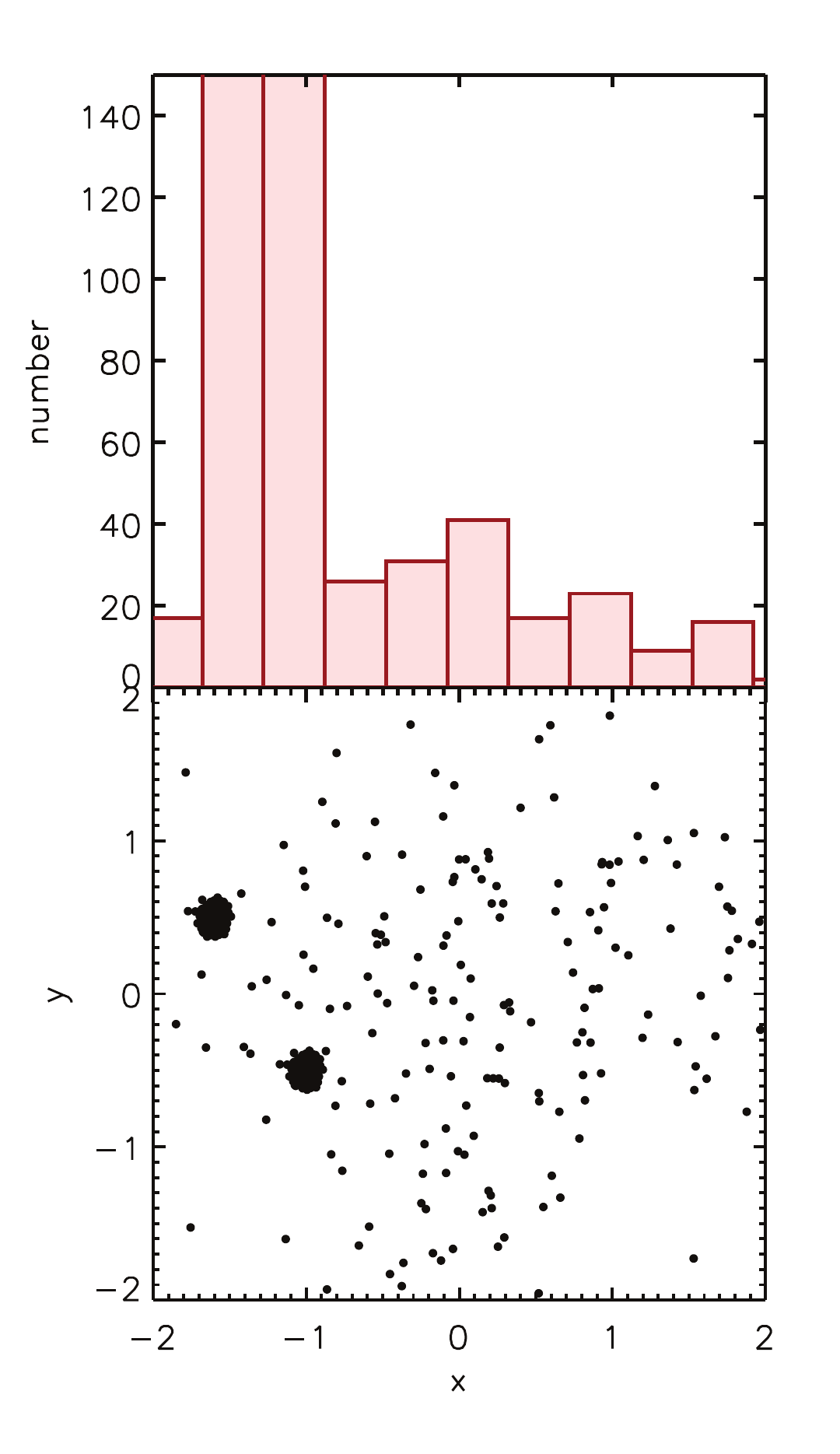} &
   \includegraphics[trim=10mm 5mm 5mm 8mm, clip,scale=0.5]{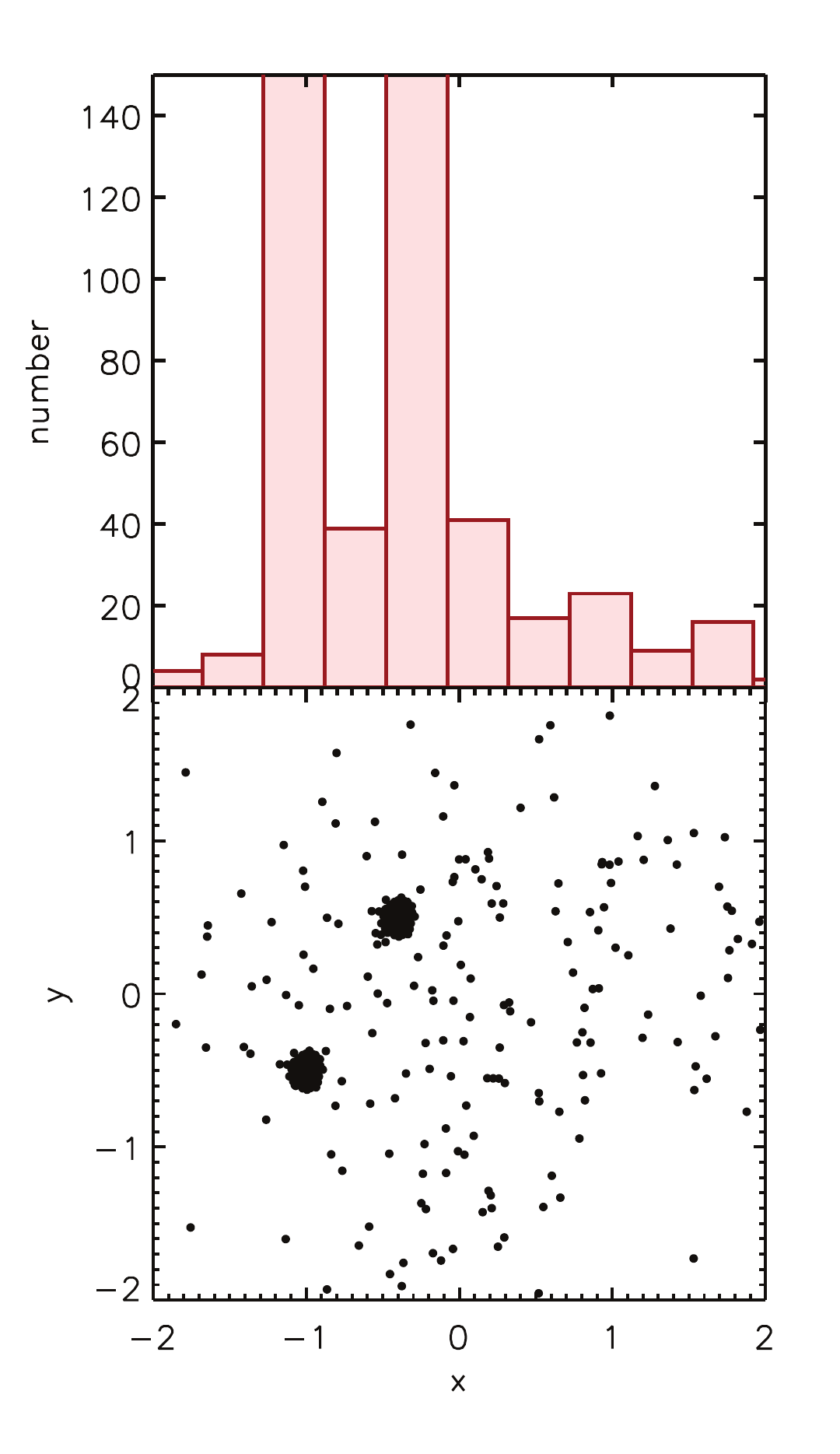}&
   \includegraphics[trim=10mm 5mm 5mm 8mm, clip,scale=0.5]{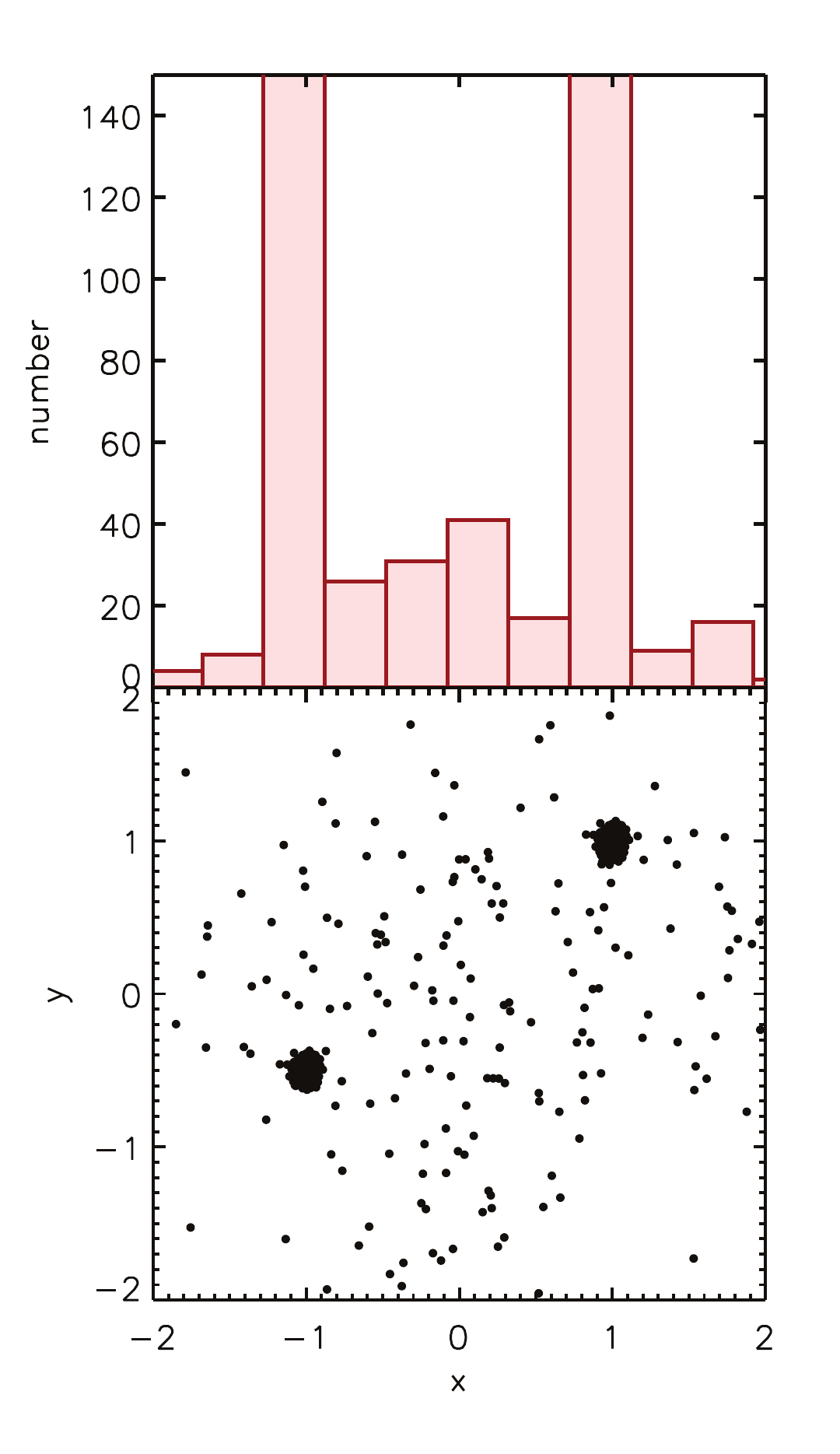}
\end{array}$
   \caption{ Three different artificial distributions are shown as examples, accompanied by the histograms of their distributions along the x axis. The background population along with one of the over-densities are similar in the three plots.  Comparing the difference in the number count of each bin alone is not indicative of the analogy  which is reflected in the similarity of adjacent bins.}
\label{illus_histo_method}
\end{figure*}

Consider the three different sample spatial distributions illustrated in Fig.\ref{illus_histo_method}, which are comparable representations of a background distribution together with two toy-clusters with different cluster distances. The histograms of their distributions along the x axis show peaks with similar height. Intuitively, we appraise the first two distributions to be more similar compared to the third one. However, comparing the Euclidean distance between the feature vectors of each pair shows the same level of similarity, with just one of the major peaks shifted to another radial bin. The Euclidean distance assumes there is no relationship between individual components (here number counts in each bin)  and does not reflect the similarity of the $i$th bin of the feature vector $\mathbf{P}$ to the \mbox{$(i+1)$-th}  neighboring bin  of the compared feature vector $\mathbf{Q}$. Therefore, the simple Euclidean distance, i. e. the difference of the count in each bin, is not appropriate for distributions which exhibit correlations of  the feature vector components. For example,  more fully occupied bins might compensate less populated neighboring bins. 

To overcome such a problem \cite{Hafner95} suggested  a method for retrieval of images based on their color histograms using a quadratic distance measure. Assuming the $N$-dimensional distributions $\mathbf{P}$ and $\mathbf{Q}$, if $\mathbf{F} =\mathbf{P} - \mathbf{Q}$, the quadratic distance can be presented as: $D_{Q}^2(\mathbf{P},\mathbf{Q})=\mathbf{F^{T}}\mathbf{A}\mathbf{F}$. The matrix $A = [a_{ij}]$ is a weighting matrix which allows for individual weighting of the cross-correlation between the bins $i$ and $j$. Therefore, we can calculate the histogram difference, $H_{d}$, by cross-matching between different components: 

$$H_{d}=D_{hist}^2(\mathbf{P},\mathbf{Q})=\mathbf{F^{T}}\mathbf{A}\mathbf{F}=\sum_{i,j}^N (p_{i}   -q_{i})(p_{j}-q_{j})   a_{ij}$$

 The histogram difference as a form of quadratic distance is successfully applied to compare color histograms of different  multimedia databases \citep[e.g.][]{Ankerst_1998}. Here we adopt the idea presented by \cite{Hafner95} to calculate the histogram difference, $H_{d}$, between various models and the observed data. Our applied weighting matrix  is constructed as:  $$a_{ij}=exp{(-\sigma  d_{ij}^2)}$$
\nnn{where $d_{ij}$ is the Euclidean distance between bins $i$ and $ j$ and $\sigma$ is a constant which determines the shape of the weighting matrix (see \cite{Hafner95} for details). This method allows us to consider the influence of neighboring bins on each other. Hence, the weighting matrix should increase the  effect of cross-correlations between the bins i and j   as a function of distance between the two bins. This objective is fulfilled by choosing lower $\sigma$ values to increase the non-diagonal entries of the weighting matrix, $A$.  On the other hand, the  weighting matrix $A = [a_{i j} ]$ should generate a quadratic distance, $H_{d}$, which is nonnegative. Such a requirement demands for higher  $\sigma$  values. In our analysis the positive semidefinite weighting matrix with relatively large non-diagonal entries is achieved with a $\sigma$ value of 12.}

The defined histogram difference, $H_{d}$, varies between 0 and $\infty$. Analogous to the Euclidean distance, the $H_{d}$ value can not be used to define the  absolute ``closeness'' or in particular ``similarity'' in this context. Nevertheless, we can use it to compare distinct models and by determining the relative ``similarity'', find the most similar model to our observed reference data set.

\subsubsection{Selection of the best-matching model }\label{sec:best_model}

 We compute the histogram difference, $H_{d}$, between the observed population of isolated massive stars in the GC and different models (see Sect. \ref{sec:models_grid}) with the aim to select the best-matching model. Table \ref{h_d} lists models with different assumptions on the age of the Quintuplet cluster as well as initial masses of stars that are currently observed as  WN, WC or OB stars. The final feature vector of the spatial distribution of massive stars in each constructed model of both the Arches and Quintuplet clusters  is compared to the observed data in the GC. Feature vectors of the 
 models are constructed as histograms of the spatial distances
 of all the stars in the two clusters with reference to the center of the Arches cluster and averaging ten random realizations of each model. The feature vector of the observed population is likewise constructed as the histogram containing the number of observed
  high-mass stars with respect to the center of the Arches cluster. To estimate the uncertainty in the histogram distance between the average model and the observations, the 10 random realizations are also individually compared to the observed distribution. The calculated standard deviation of the resulting $H_{d}$ values are reported for each model in Table \ref{h_d}. Since we are only interested  in the population of drifted or ejected stars, sources that lie inside the tidal radius ($\sim 1.6\,\mathrm{pc}$; \citealt{habibi}) of the Arches, Quintuplet, and Nuclear clusters are excluded from the data as well as the models. The bin size is chosen to be $3$ pc to avoid random fluctuations. Our experiment shows that the final calculated $H_{d}$ value is not very sensitive to the bin size since our method accounts for the impact of neighboring bins on each particular bin.  The model with an age of 5  Myr for the Quintuplet cluster and an initial mass of $40\, M_{\odot}$ for massive stars is found to have the lowest $H_{d}$ value and is the most similar model to the observed distribution. We use this best-matching model for the comparison to the observations and further analysis.

 \begin{table*}[t]
\caption{Calculated histogram difference, $H_{d}$, between the observed isolated massive sources by \cite{Mauerhan2010_main} and different models. Compared models assume an age of 4, 4.5 and, 5 Myr for the Quintuplet cluster and 2.5 Myr for the Arches. The initial mass for the massive stars is considered to be 20, 40, and $60\,M_{\odot}$ in these models. The model which assumes an age of 5  Myr for the Quintuplet cluster and an initial mass of $40\,M_{\odot}$  for WR progenitors is found to have the lowest $H_{d}$ and is the most similar model to the observed distribution. The standard deviations for  the $H_{d}$ values are derived by individual comparison of the 10 random realizations of each model.}             
\label{h_d}      
\centering          
\begin{tabular}{c|c c c c c c c  }     
\hline
&&&$H_{d}$&\\
\hline     
\\  
&WR-L=20$M_{\odot}$ &  StdDev & WR-L=40 $M_{\odot}$ & StdDev & WR-L=60 $M_{\odot}$& StdDev\\ 
\hline                 
$Q_{age}$=4 Myr & 17.4 &$\sigma_{4,20}=4.38$& 4.48 &$\sigma_{4,40}=1.43$& 6.38&$\sigma_{4,60}=1.03$ \\   
$Q_{age}$=4.5 Myr &  26.55 &$\sigma_{4.5,20}=6.93$& 5.96 &$\sigma_{4.5,40}=1.15$& 5.63&$\sigma_{4.5,60}=0.88$ \\  
$Q_{age}$=5.0 Myr &  11.69 &$\sigma_{5,20}=1.79$&  \bf{3.784} &$\sigma_{5,40}=0.75$& 7.25&$\sigma_{5,60}=0.81$ \\  
\hline                  
\end{tabular}
\end{table*}

\begin{table*}[t]
\caption{Positions and spectral types of the observed isolated massive sources in the GC region within the area covered by the Pa$\alpha$ survey. From left to right the columns are: Sequential ID for stars, R.A., Dec., spectral type, and the reference study (a). \nnn{The last two columns of the table present the Poisson distribution, ${P(n)=\lambda^{n} e^{-\lambda}/n!}$, of observing  one star, $P(1)$, and one or more stars, ${P(n\geq 1)=\sum_{i=1}^{\infty} P(n)}$, in the assigned Voronoi cell of each particular star on the density map of the observed isolated massive stars illustrated in Fig. \ref{d_vor_map}. It is important to notice that the maximum value of the Poisson probability of observing one star, $P(1)=\lambda e^{-\lambda}$, is $\sim 37\%$.}}             
\label{source_list}      
\centering          
\begin{tabular}{c c c c c |c c } 
\hline\hline       

ID & $ R.A.$ & $ Dec.$  & Spectral  & references  & P(1) & P($\geq$1) \\ 
 & (deg, J2000)  &  (deg, J2000) & type &   \\ 

\hline                    
1 & 266.2797  & -29.199789 & WC & 1  & 0.000 & 0.000\\
2 & 266.34123 & -29.199841 & WC & 1  & 0.335 & 0.777\\
3 & 266.29086 & -29.236897 & WC & 1  & 0.268 & 0.330\\
4 & 266.35067 & -29.01608  & OB & 1  & 0.222 & 0.259\\
5 & 266.38129 & -28.954669 & OB & 1  & 0.222 & 0.259\\
6 & 266.38549 & -29.082757 & WC & 1  & 0.311 & 0.817\\
7 & 266.42203 & -28.863311 & OIf & 1 & 0.222 & 0.259\\
8 & 266.42639 & -28.879828 & OB & 1  & 0.090 & 0.095\\
9 & 266.47251 & -28.827035 & WN & 1  &  0.193 & 0.926\\
10 & 266.50699 & -28.920983 & OI & 1 & 0.164 & 0.181 \\
11 & 266.51091 & -28.903941 & WC & 1 & 0.359 & 0.551\\
12 & 266.54181 & -28.925694 & WN & 1 & 0.090 & 0.095 \\
13 & 266.57324 & -28.884391 & WN & 1 & 0.090 & 0.095\\
14 & 266.59932 & -28.803129 & WN & 1 & 0.002 & 1.000 \\
15 & 266.54652 & -28.818221 & WN & 1 & 0.031 & 0.994\\
16 & 266.4607  & -28.957282 & WC & 1 & 0.222 & 0.259\\
17 & 266.52344 & -28.858866 & LBV & 1 & 0.368 & 0.632\\
18 & 266.26195 & -29.14986  & OB & 2 & 0.000 & 0.000\\
19 & 266.28733 & -29.20495  & WN & 2 & 0.000 & 0.000\\
20 & 266.31744 & -29.0543   & WN & 2,4 & 0.000 & 0.000\\
21 & 266.31969 & -28.97364  & WN & 2 & 0.000 & 0.000\\
22 & 266.32974 & -29.05609  & WC & 2 & 0.090 & 0.095 \\
23 & 266.34453 & -28.97895  & WN & 2 &  0.090 & 0.095\\
24 & 266.38652 & -28.93797  & OI & 2 &  0.329 & 0.451\\
25 & 266.40056 & -28.94405  & WN & 2 &  0.000 & 0.000\\
26 & 266.40538 & -28.89827  & OB & 2 &  0.354 & 0.727\\
27 & 266.48067 & -28.85738  & WN & 2,3  & 0.090 & 0.095\\
28 & 266.57119 & -28.85871  & Of & 2,3  & 0.222 & 0.259\\
29 & 266.5743  & -28.83541  & WC & 2 &  0.366 & 0.667\\
30 & 266.46089 & -28.98879  & WN & 2,4,5  & 0.359 & 0.551\\
31 & 266.45185 & -28.834709 & WN & 5  & 0.231 & 0.900\\
32 & 266.42696 & -28.881472 & WC & 6  & 0.090 & 0.095\\
33 & 266.49075 & -28.912806 & WC & 6  & 0.268 & 0.330\\
34 & 266.56456 & -28.838492 & WC & 7  & 0.361 & 0.699\\
35 & 266.41381 & -28.88923 & OB & 4,5 & 0.222 & 0.259\\
\hline                  
\end{tabular}

\begin{list}{}{}

\item[(a)] References: (1) \cite{Mauerhan2010_main}; (2) \cite{Mauerhan_matching}; (3) \cite{2007ApJ...662..574M}; (4) \cite{2006ApJ...638..183M}; (5)  \cite{1999ApJ...510..747C}; (6) \cite{2003A&A...408..153H}; (7) \cite{Figer1999}
\end{list}
\end{table*}

\section{Results}
\label{sec: results}

The Arches cluster moves rapidly through the high-density environment in the GC (e.g. \citet{2008ApJ...675.1278S}). As the cluster moves along its orbit, two body relaxation in the cluster accelerates some stars to exceed the escape velocity, which is variable and determined by the tidal force. These stars either take over the cluster center or fall behind depending on their differential velocities. This known phenomenon populates the two extended tidal arms. In this section, we present the comparison of the massive stars outside the Arches and Quintuplet clusters in the best-matching model, which mostly belong to the tidal arms,  with the observed massive sources in isolation  \citep{Mauerhan2010_main}. In Sect. \ref{sp_dist}, simulations are compared to the data in terms of the number of isolated sources and their spatial distribution. In Sect. \ref{velocity_anal}, the velocity variation along the tidal structures will be analyzed to obtain a general picture of dynamical evolution of the cluster as it moves along its orbit.
\subsection{The spatial distribution of drifted and observed high-mass stars}
\label{sp_dist}

\begin{figure*}[t]
\includegraphics[scale=0.6]{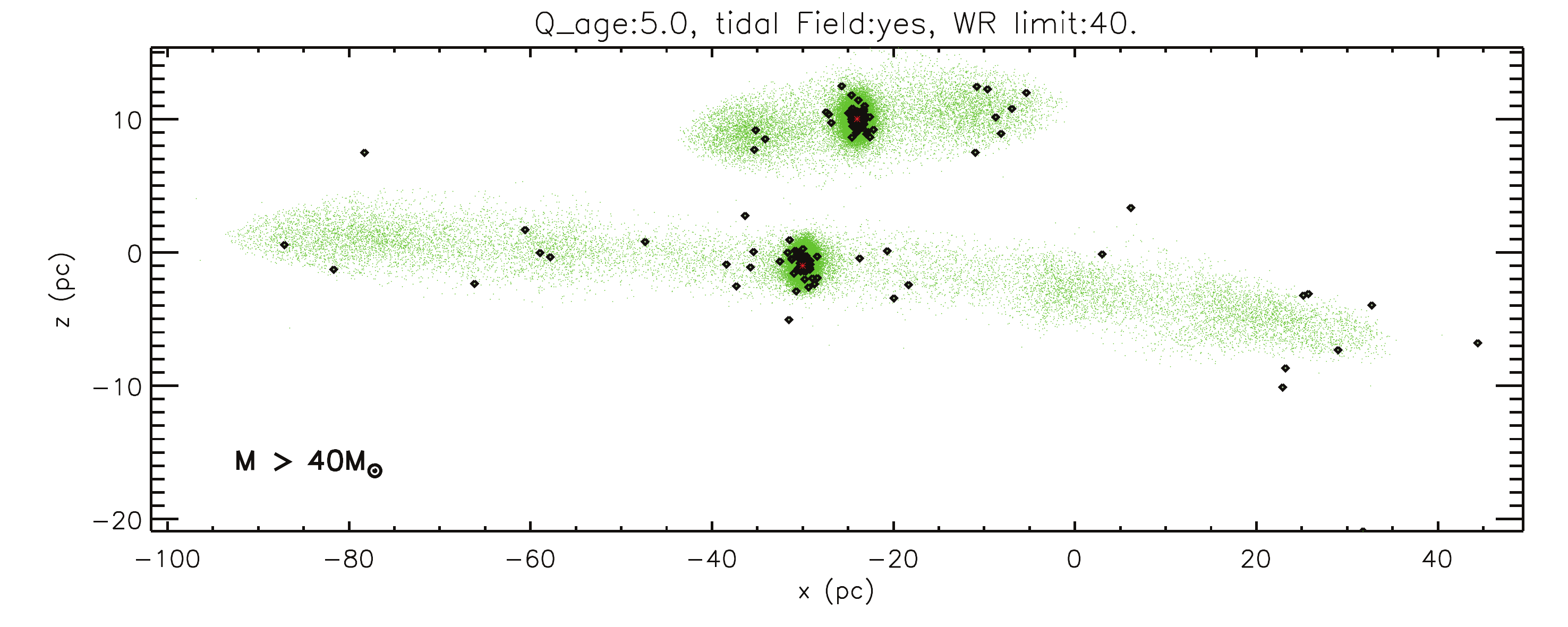}
\caption{ One realization of the best-matching model is projected on the plane of the sky. Green dots represent the cluster members. Sources with initial masses exceeding $40\, M_{\odot}$ are illustrated by filled diamonds. The x-axis is along the Galactic plane and the z-axis is towards the Galactic north pole.  }
\label{cluster_look}
\end{figure*}

The best-matching model of massive sources outside the Arches and Quintuplet clusters assumes an age of $5 Myr$ for the Quintuplet cluster and $M>40\,M_{\odot}$ for WR progenitors. Figure \ref{cluster_look} demonstrates the projection of one of the realizations of the best-matching model on the plane of the sky. In this model, each tidal arm of the Arches cluster extends out to $20\,\mathrm{pc}$ along the Galactic plane, whereas tidal arms of the Quintuplet cluster stretch out to $65\,\mathrm{pc}$. Massive stars in both clusters, $M > 40 M\odot$, are mostly concentrated around the cluster center. Mass segregation of the Arches cluster has been observationally confirmed since the observed slope of the mass function increases with radius (\citealt{Espinoza}, \citealt{habibi}). Observations of the  high-mass part of the mass spectrum of the Arches cluster ($M>\sim 10 M_{\odot}$)  out to its tidal radius revealed a depletion of massive stars in the cluster outskirts \citep{habibi}. In our previous study, we derived a present-day mass function  slope of $-3.21 \pm 0.30$ in the outer annulus ($0.4 < r < 1.5\,\mathrm{pc}$) as compared to a flat slope of $-1.50 \pm 0.35$ in the core ($r < 0.2\,\mathrm{pc}$), where the Salpeter slope is $-2.3$. The Quintuplet cluster exhibits about 100 times lower density in the core compared to the Arches cluster (\citealt{Figer2008}). The  present-day mass function of the Quintuplet cluster is also characterized by a flat slope, $ -1.68 \pm 0.1$ \citep{benjamin}. Apart from the high-mass stars in the cluster centers, Fig. \ref{cluster_look} illustrates that, while the tidal arms are mostly constructed from low and intermediate mass stars, in the models the majority of massive sources outside the cluster centers are also located along the tidal arms. 

\begin{figure*}[t]
   \includegraphics[scale=0.6]{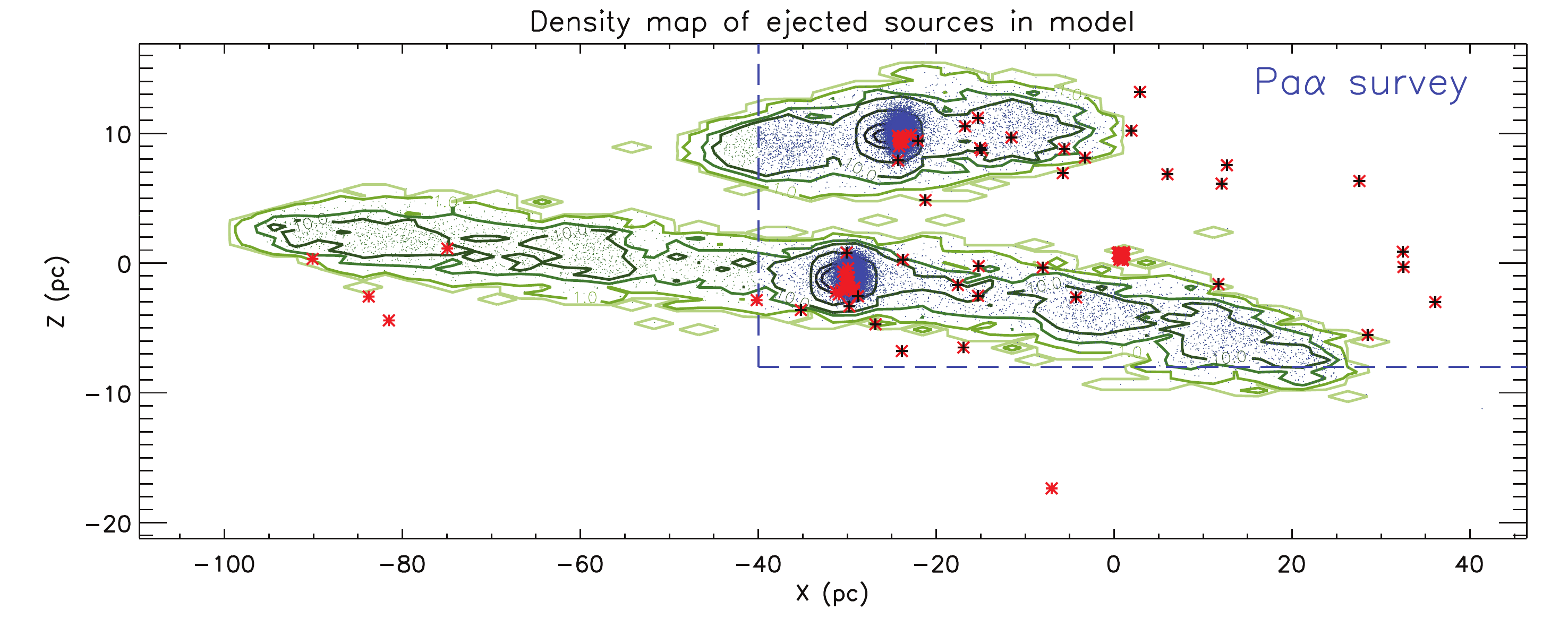}

   \caption{ The contour density map for one of the realizations of the best-matching model, \nnn{with an age of 2.5 and 5  Myr for the Arches and Quintuplet clusters ,respectively, and an initial mass of $40\, M_{\odot}$ for massive stars}, is presented with different shades of green lines. Darker colors indicate higher densities. The isolated massive stars observed by \cite{Mauerhan2010_main} are marked with red asterisks. \nnn{The  dashed lines approximately represent} the  border of the Pa$\alpha$ survey. This line separates the region  where we perform our  comparison between the model and the data.\nnn{ We include some of the observed sources outside the  Pa$\alpha$ survey area for illustration. Only the observed sources, shown by red asterisks, which are marked  with black  crosses are used to compare observations with the models.} Blue (green) dots represent the simulated cluster members which are included (excluded) for the comparison analysis. The x-axis is along the Galactic plane and the z-axis is towards the Galactic north pole.  }
\label{dens_map}
\end{figure*}

A contour density map of the same realization of the best-matching model is shown in Figure \ref{dens_map} along with the population of observed isolated massive stars. These sources fill two stripes along the Galactic plane with a distinct gap in the separation space between the two clusters along the direction of the galactic poles. The observed configuration is closely reproduced in the model by the location of the tidal arms of the two clusters. In Fig.\ref{dens_map}, the Nuclear cluster is located at the origin, (0,0). As the minor contribution of the  Nuclear cluster was justified in Sect. \ref{sec:models_grid}, our models \nnn{do not include} the Nuclear cluster.

\begin{figure*} 
\centering $
\begin{array}{cc}
   \includegraphics[trim=10mm 5mm 5mm 8mm, clip,scale=0.5]{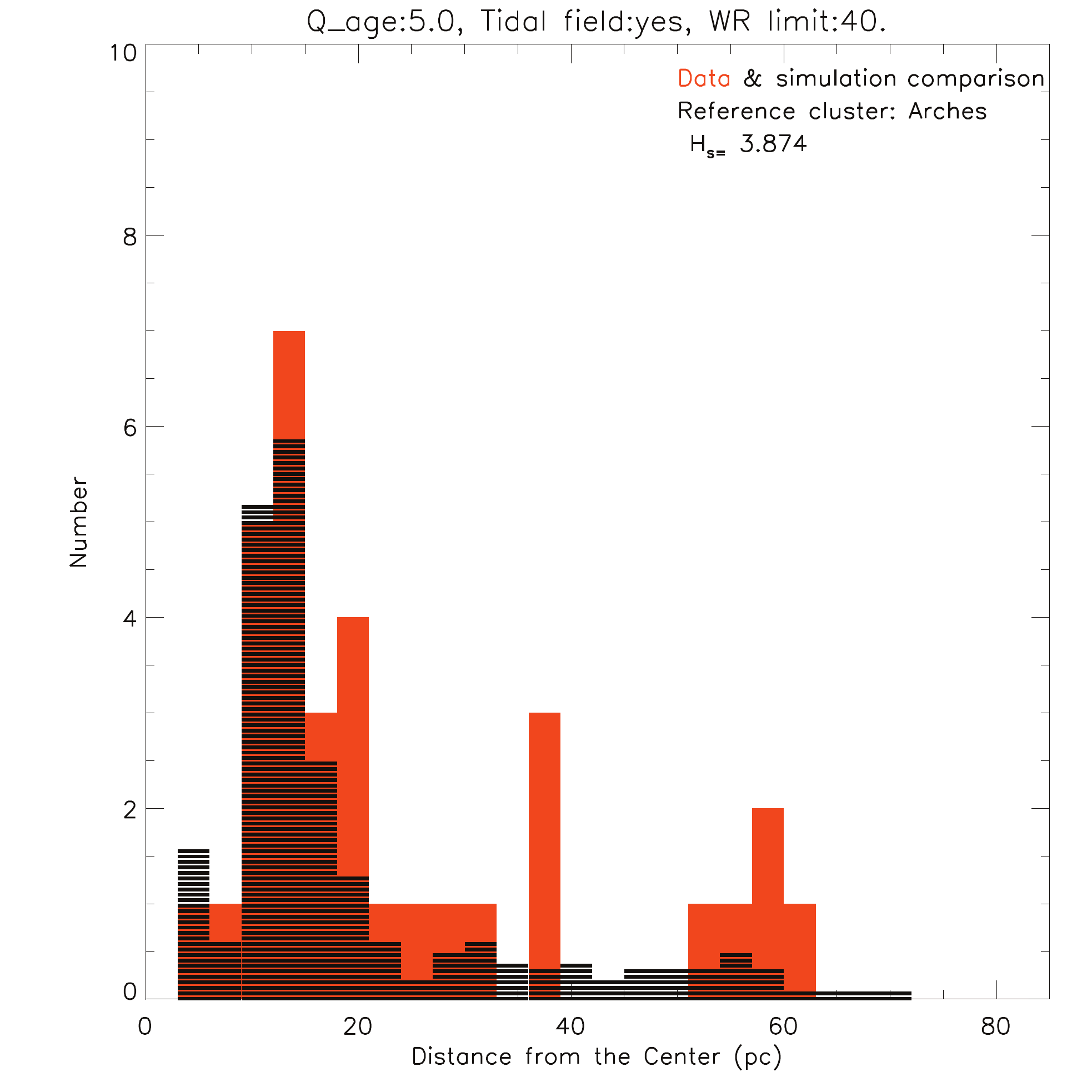} &
   \includegraphics[trim=10mm 5mm 5mm 8mm, clip,scale=0.5]{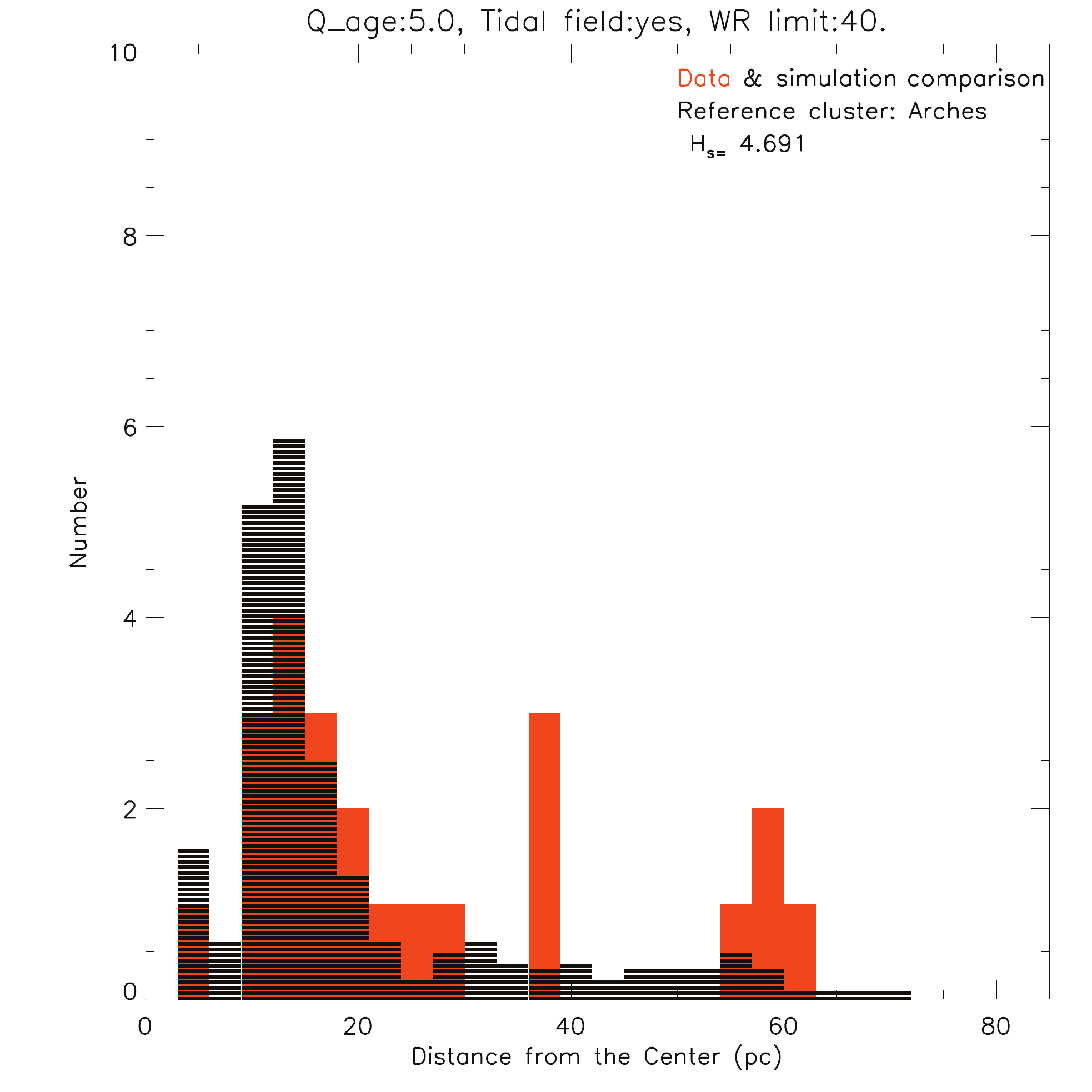}
\end{array}$

   \caption{Line filled histograms depict the average distribution of the massive sources, $M>40\, M_{\odot}$, in the ten random realizations of the best-matching model. Stars within the central $1.6\,\mathrm{pc}$ of the simulated Arches and Quintuplet clusters are excluded. Red filled histograms illustrate the distribution of the observed massive sources. Distances are calculated with reference to the center of the Arches cluster. Observed sources which lie in the central $1.6\,\mathrm{pc}$ of the three clusters, i. e. Arches, Quintuplet, and Nuclear cluster, are excluded from these histograms. The calculated histogram difference, $H_{d}$, measures the similarity of the two distributions. Left: The comparison is based on the full list of WR stars and OB supergiants presented  by \cite{Mauerhan2010_main}. Right: OB super giants are excluded from the observed catalog and the comparison is based on the population of WRs in the region.   }
\label{histo}
\end{figure*}

\nnn{A group of observed isolated massive stars is located outside the Pa$\alpha$ survey area around $x \sim-80\,\mathrm{pc}$ in Figure \ref{dens_map}. These stars along with more X-ray emitting sources are thought to be associated with the Sagittarius B molecular cloud complex \citep{Mauerhan_matching}. Figure \ref{dens_map} illustrates that the tidal arms of the Quintuplet cluster might  extend out to the Sagittarius B region and therefore the observed evolved massive stars in this region might in fact originate from the tidally drifted structures of the Quintuplet cluster. }

The predicted number of massive stars based on the full set of ten different random realizations of each model is presented in Table \ref{number_massive}. The models with an  initial mass of $20\, M_{\odot}$ for WR progenitors overpredict the expected number of WR stars which should have been detected in the GC region, while $M>60\,M_{\odot}$ provides too few sources to represent the observed population. As Table \ref{number_massive} shows, $M>40\,M_{\odot}$ is likely the most realistic initial mass of the progenitor sources that evolve to the observed isolated high-mass population today. These models predict 22-30 isolated high-mass stars which cover a significant fraction of the 35 observed massive stars in the Pa$\alpha$ survey region outside the three clusters (outside their approximate tidal radius) in the GC region.
\begin{table*}[t]
\caption{ In the Pa$\alpha$ survey region 35 massive stars are observed outside the three clusters (outside their approximate tidal radius) in the GC region. Among these massive stars are 24 Wolf-Rayet stars. Below the predicted number of massive stars of each model is presented. }             
\label{number_massive}   
\centering          
\begin{tabular}{c|c c c c c c c  }     
\hline
& &\multicolumn{3}{c}{$N_{massive}$ outside the clusters}  \\
\hline       
\\

&WR-L=20$M_{\odot}$ &  StdDev & WR-L=40 $M_{\odot}$ & StdDev & WR-L=60 $M_{\odot}$& StdDev\\ 

\hline                 
$Q_{age}$=4 Myr & 62&$\sigma_{4,20}=8$ & \bf{22}&$\sigma_{4,40}=5$ & 11&$\sigma_{4,60}=4$ \\   
$Q_{age}$=4.5 Myr &  80 &$\sigma_{4.5,20}=12$& \bf{30} &$\sigma_{4.5,40}=4$& 13&$\sigma_{4.5,60}=2$ \\  
$Q_{age}$=5.0 Myr &  68 &$\sigma_{5,20}=8$&  \bf{26}&$\sigma_{5,40}=3$ & 10&$\sigma_{5,60}=2$ \\  
\hline                  
\end{tabular}
\end{table*}

\subsection{\nnn{Comparing the spatial distributions}}

Table \ref{number_massive} allows to perform a comparison between the number of massive stars outside the clusters in the models and the observations, independent of their spatial distribution. To compare the spatial distribution of drifted sources in the models with the data, we use two different methods. In Sect. \ref{voronoi},  we present the predicted density map of massive stars based on our best-matching model. In Sect.\ref{histo_comp},  the feature vector of simulated sources is constructed on the full set of ten different random realizations of the best-matching model (see Sect. \ref{sec:best_model}) and is compared with the data. 

\begin{figure*}

\centering $
\begin{array}{cc}

\includegraphics[trim=0mm 0mm 10mm 0mm, clip,width=110mm]{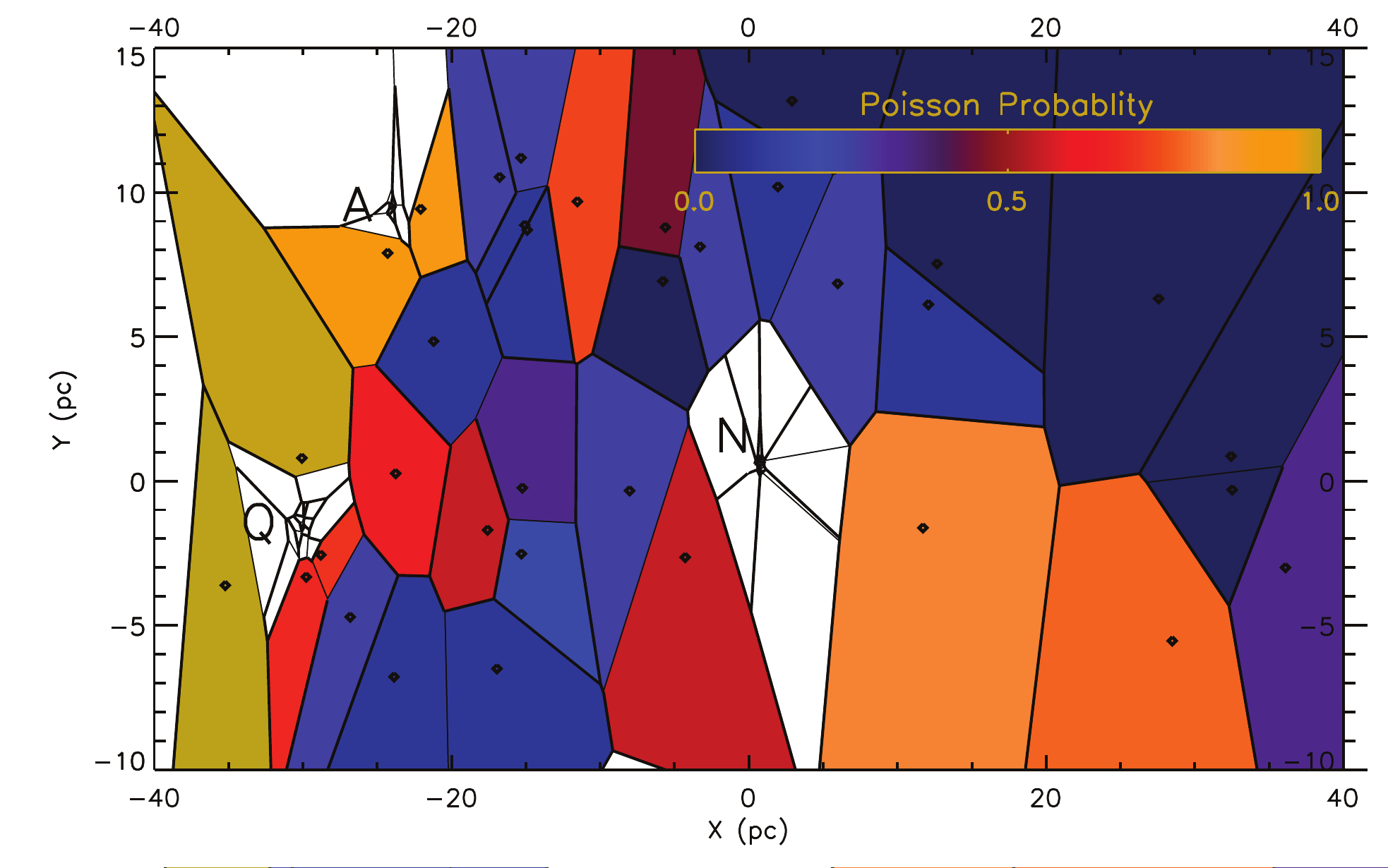}&
\includegraphics[trim=0mm 5mm 5mm 5mm, clip,width=65mm]{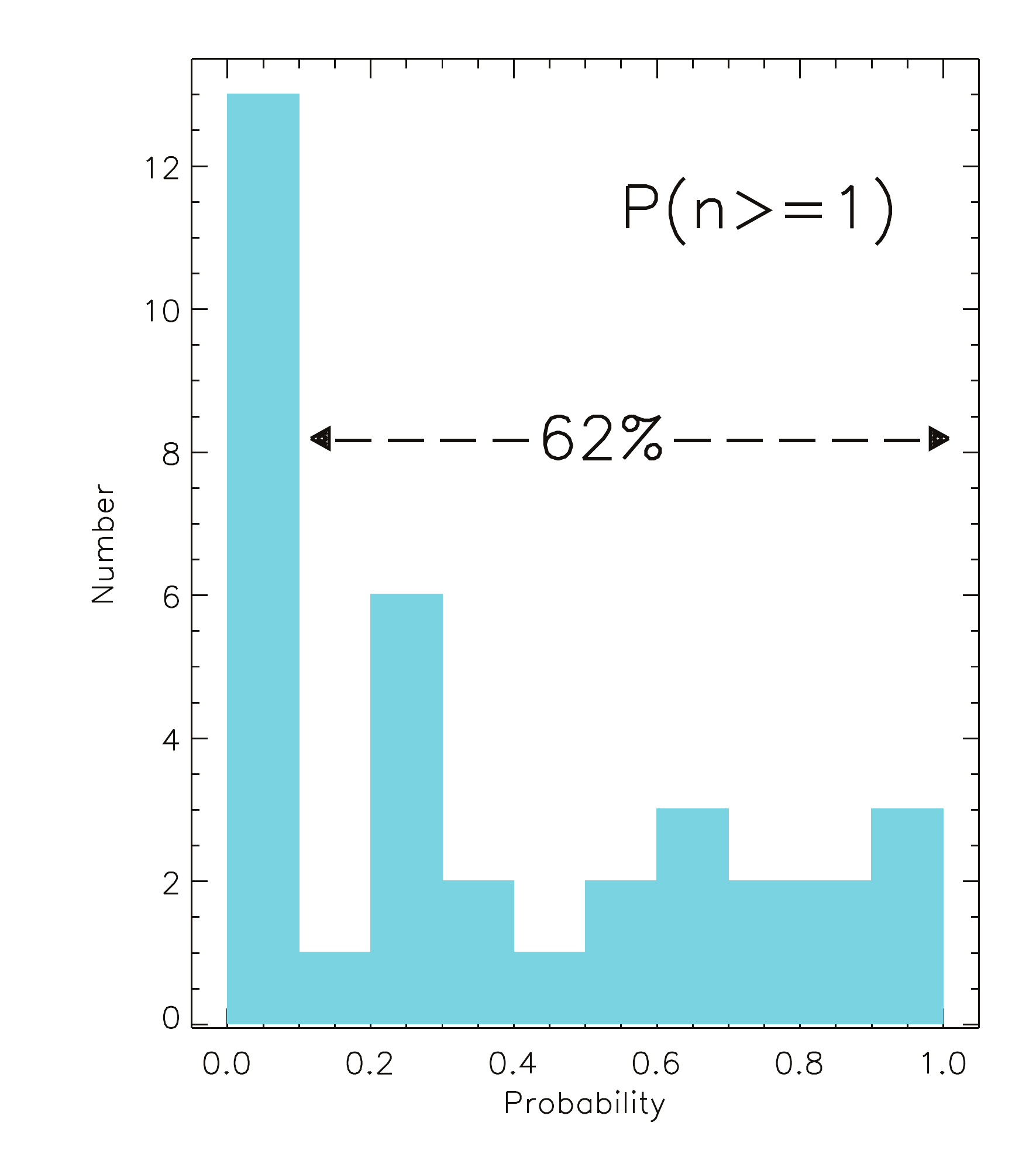} 
\end{array}$
\caption{\nnn{Left: The density map of the observed massive stars \citep{Mauerhan2010_main} is derived using Voronoi diagrams. The two stripes of polygons formed beside the clusters and along the x axis confirm the preferred orientation of the observed massive stars. The Voronoi cells are colored according to their calculated Poisson probability of observing one or more stars in each cell, $P(n\geq1)$, based on our best model. The color-bar assigns colors to the calculated probabilities. White regions contain the cluster central members and are discarded from the probability calculations. Right: The distribution of calculated Poisson probabilities, $P(n\geq1)$, is shown. Our model reproduces at least one star in ten random realizations, $P(n\geq 1)>10\%$, for 62\% of the observed isolated massive stars. Please note that the Voronoi cells appear skewed due to different scales for the x and y axis. }}
\label{d_vor_map}
\end{figure*}

\nnn{
\subsubsection{The density map}\label{voronoi}

We create a density map of the observed massive stars using Voronoi diagrams. Voronoi tessellations define a neighborhood for each star, the Voronoi cell, on the plane of the sky. Each cell contains one observed star and the set of points on the plane of the sky  which are closest to the generating star of the cell. Employing Voronoi diagrams allows us to see the hidden spatial structure in the spread of the points. For example, a preferred orientation in the distribution of points reflects as oriented polygons \citep{vor_book}.

Fig \ref{d_vor_map} illustrates the resulting density map. The density map shows two stripes of polygons formed beside the clusters and along the x axis, which confirm the preferred orientation of the observed massive stars along the clusters' tidal tails. The colors of the cells in Fig. \ref{d_vor_map} are assigned by the calculated probability of observing one or more stars in each particular cell, according to our best-matching model. In order to calculate the probabilities we count the average number of predicted modeled massive stars, $M> 40\,M_{\odot}$, in each Voronoi cell, for all the available random realizations of the model. The corresponding probabilities are accordingly calculated assuming a Poisson distribution, ${P(n\geq 1)=\sum_{i=1}^{\infty} P(n)}$. The Poisson probability, ${P(n)=\lambda^{n} e^{-\lambda}/n!}$, yields the probability of observing $n$ stars where the model predicts the mean value, $\lambda$, for the number of stars in each Voronoi cell. 

 Fig \ref{d_vor_map} (Left) illustrates that the area with a high probability of observing one or more stars is located at a distance of $\sim 20$ pc from the cluster centers. The regions which contain the cluster central members are discarded from the probability calculations (see white regions in Fig. \ref{d_vor_map}). The model predicts relatively high probabilities, $P(n\geq 1)$, along the tidal arms and specifically at the end of the arms. The end of the tidal arms are located at $(x\sim 0,y\sim 10)$ pc for the Arches cluster, and $(x\sim 35,y \sim -7)$ pc for the Quintuplet cluster. Fig \ref{d_vor_map} (Right) shows the distribution of calculated probabilities, ${P(n\geq 1)}$. For  62\% of the observed isolated massive stars at least one of the ten random realizations of our model predicts  a star that can explain the observed star. This number increases to 72\% when we only consider the Voronoi cells within the central 20 pc from the center of the Arches cluster. 

The areas of low probabilities are located in the upper right of  Fig. \ref{d_vor_map},  $10 \lesssim x \lesssim 40,\, 0 \lesssim y \lesssim 10$ pc, and also in the region which lies between the two clusters, $3<y<7$ pc. These are the sources which, according to our models, are less probable to originate from the tidal arms of the Arches and Quintuplet clusters. In Sect. \ref{alternatives} we discuss the possible alternative and parallel scenarios that may explain the origin of these sources. Computed probabilities to observe exactly one star, ${P(n=1)}$, and also to observe one or more stars, $P(n \geqslant 1)$, are presented in Table \ref{source_list}. It is important to notice that the maximum value of the Poisson probability for observing one star, $P(1)=\lambda e^{-\lambda}$, is $\sim 37\%$. }

\subsubsection{Comparing feature vectors}\label{histo_comp}

Figure \ref{histo} illustrates histograms of distances of all the massive sources with reference to the center of the Arches cluster.  It shows that our models produce drifted massive sources at a distance of up to $70\,\mathrm{pc}$ from the Arches center.  This result implies that for an observed isolated massive star at a distance of up to $70\,\mathrm{pc}$ from the Arches center, the possibility of being a run away/drifted star especially from the Quintuplet cluster can not be excluded.  The provided list by \cite{Mauerhan2010_main} includes 11 WC and 13 WN stars together with 1 LBV and 10 OB supergiants  outside the $1.6\,\mathrm{pc}$ radius of the three clusters in the Pa$\alpha$ survey region. The comparison of their full list in the Pa$\alpha$ survey area to the model is shown in Fig. \ref{histo}, left. The best-matching model reproduces $80\%$ of the known isolated massive stars out to a distance of $21\,\mathrm{pc}$ and  $67\%$ of the observed population out to $80 \,\mathrm{pc}$ from the center of the Arches cluster.  However, since the population of OB supergiants is not complete in this catalog, we excluded all the OB stars from the original observed list and repeated the comparison. This comparison shows that our model produces 20\% more massive sources as compared to the observed population of WN, WC and LBV stars in the GC region out to a distance of $21\,\mathrm{pc}$ from the center of the Arches cluster  (see Fig \ref{histo}, right). This over-prediction can be a result of the fact that we can only choose an initial stellar mass. We have no means to distinguish the spectral types of the stars in our models. Hence, stars which are in the supergiant phase or even main sequence phase will always be partially included. It can be also due to a missing population of WC stars. The Pa$\alpha$ line strength of both WN and WC stars span a similar range of $1.2<F187/F190N<2.9$ \citep{Mauerhan2010_main}. The bright $K_{s}$ counterparts of WN stars guarantees a near completeness of the WN sample in the GC. However, due to the faint end of the $K_{s}$ brightness distribution of observed WC stars their sample is probably not complete (see \cite{Mauerhan2010_main} for details).

\subsection{\nnn{Alternative scenarios}}\label{alternatives}

 In both histograms of Fig. \ref{histo}, the major peak in the observed distribution of isolated massive stars at $\sim 15\,\mathrm{pc}$ is well reproduced by the best-matching model. Nevertheless, the two minor peaks at distances of $\sim 40\,\mathrm{pc}$ and $\sim 60\,\mathrm{pc}$ from the Arches center are absent in the model. Out of the 35 observed massive sources, 27 stars are located close to the modeled tidal arms of either of the two clusters. The remaining 8 sources are observed outside the modeled arms. These sources which are not reproduced with our models can have different origins: Many high mass stars are expected to be in binaries with high-mass companions ($\sim70\%$ of all O stars of which $\sim40\%$ are expected to have evolved with strong interactions; \citealt{2012Sci...337..444S}). The observed sources outside the tidal structures can originate from tight binary systems in which the primary has already exploded as a supernova (\cite{1961BAN....15..265B}; \cite{1991AJ....102..333S}). Therefore, the secondary has received a kick and is ejected at a random direction and at a high relative velocity compared to the cluster orbital velocity. Such ejecta are expected to have a high velocity compared to the velocity of the cluster center and can possibly explain the detected population in between the modeled tidal structures as well as the single detected massive star at z=-18 pc (see Fig. \ref{dens_map}). This process is particularly expected from some of the most massive stars in the Quintuplet cluster at the suggested age of 4 to 5 Myr. Another possibility for their origin is that they are formed independently of the current observed clusters.

  A possible Orion Nebula  (ONC)-type host star cluster which owns $\sim 1$ massive source of $\sim 35\, M_{\odot}$ can not be detected in the GC region with any of the current wide-field surveys.  The density of such a star cluster is too low and intermediate-mass sources ($M<10\, M_{\odot}$) are too faint to stand out among the dense, crowding-limited GC field. These clusters are expected to dissolve even faster than the young sturburst clusters. The survival life-time of less than 10 Myr suggests that they could appear too dispersed to be detected as a compact entity, such as e.g. the Arches cluster (for discussion on dissolution time and detectability of the clusters in the GC see \citealt{2001ApJ...546L.101P}). The other scenario is that these sources are formed in isolation from the dense molecular clouds observed in the central molecular zone \citep{2012ApJ...746..117L}. The last two scenarios are currently hard to distinguish.
 
The discrepancies between the model and the data also can originate from limitations of our models. 
In the absence of \nnn{precise} 3D velocity measurements for the Quintuplet cluster\footnote{\nnn{The preliminary analysis of the proper motion data of the Quintuplet cluster yields  $172 \pm 50, \mathrm{km s^{-1}}$ \citep{2011sca..conf..304S} which is surprisingly similar to the measured value of $172 \pm 15\, \mathrm{km s^{-1}}$ for the Arches cluster \citep{2012ApJ...751..132C}. Given the close projected distance between the Arches and Quintuplet clusters in the GC region, we hence use the Arches orbit to simulate the Quintuplet cluster.}}, our models are assuming that the Quintuplet cluster is an older representation of the Arches cluster. Although this assumption is suggested based on the observed density and age of the Quintuplet cluster, \nnn{a more accurate} measurement of the 3D velocity of the Quintuplet cluster, will eventually help to choose the best matching orbit for the Quintuplet. Employing different orbits can shift the position of the tidal tails on the plane of the sky. We expect that our experiment with assuming three different ages for the Quintuplet cluster partly covers the effect of uncertain orbit of the Quintuplet cluster. The other limitation of the models is the lack of primordial binaries. Although wide binaries form automatically in the models in the early stages of cluster evolution, the population of tight initial binaries is missing in these simulations. These binaries play an important role  in dynamical three
or four-body encounters which produce ejected stars (e.g.  \cite{1967BOTT....4...86P}; \citealt{1986ApJS...61..419G}). Adding primordial binaries to these models will produce more massive sources outside the tidal radius of the clusters and strengthen the similarity of the models with the data. These sources are less likely to be spatially confined to the tidal arms of the cluster, yet the effect of the GC tidal field on these high-velocity ejected sources needs a thorough investigation in future studies.

In summary, the dynamical model of the Arches and Quintuplet clusters, assuming an initial mass of $40\,M_{\odot}$ for the WR star progenitors and a Quintuplet age of $5\,\mathrm{Myr}$, can explain up to 80\% of the observed isolated population of massive stars.

 \subsection{The velocity distribution of drifted cluster members}
  \label{velocity_anal}
  
\begin{figure*}

\centering $
\begin{array}{cc}
\includegraphics[trim=10mm 5mm 5mm 5mm, clip,width=170mm]{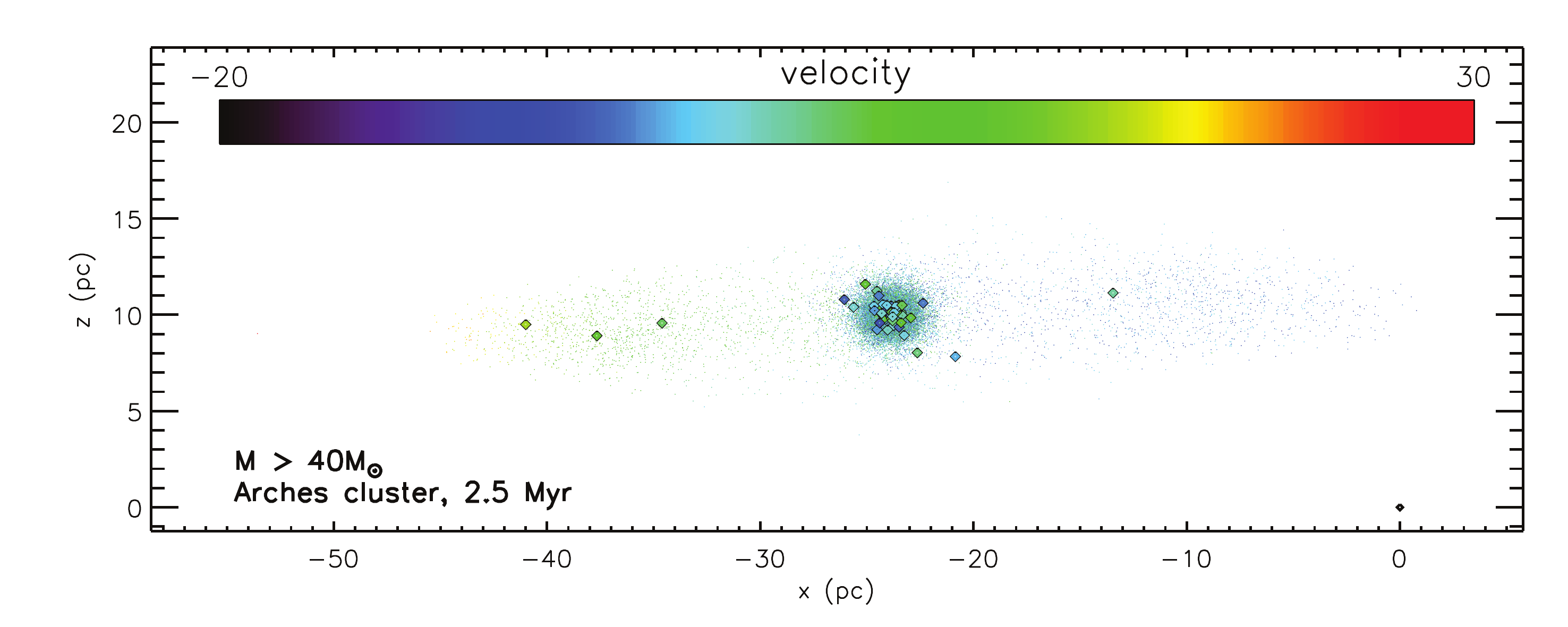} \\
\includegraphics[trim=10mm 5mm 5mm 8mm, clip,width=170mm]{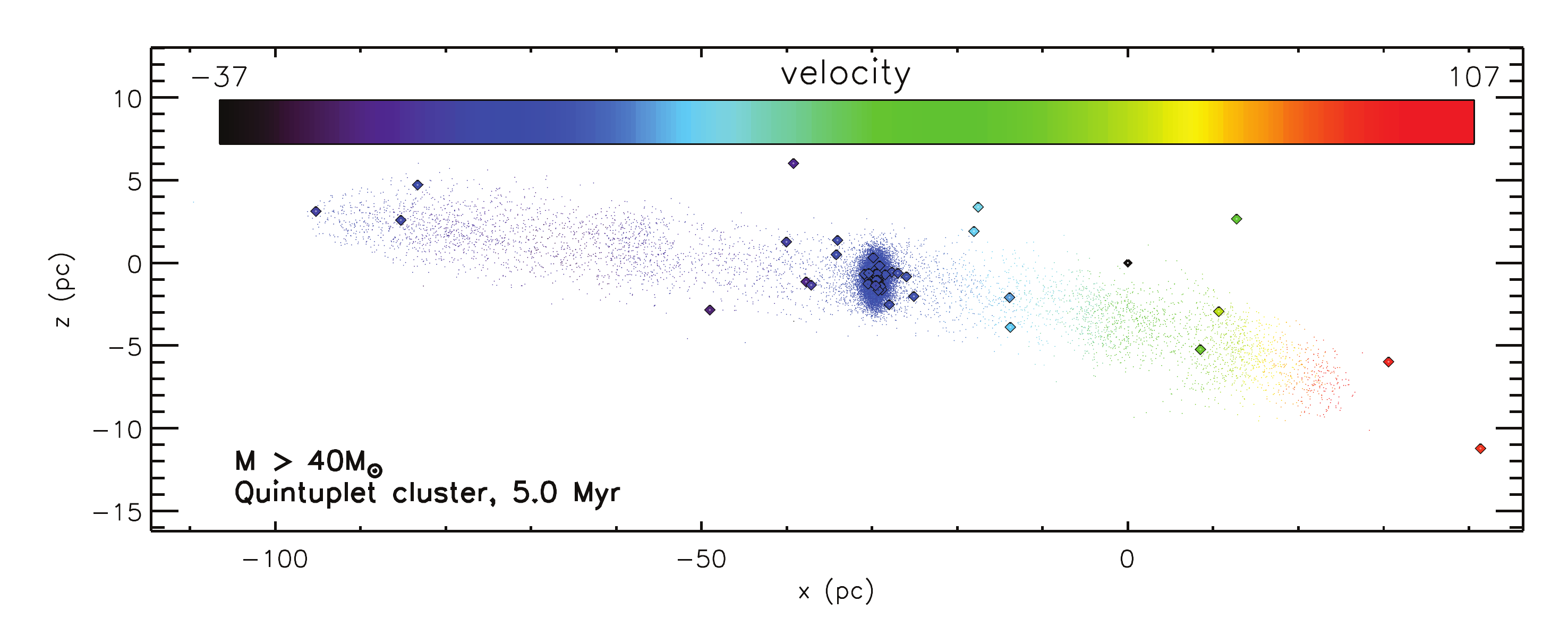}
\end{array}$
\caption{Velocity variation within the tidal arms for one realization of the simulated cluster at an age of 2.5 Myr (up) and 5 Myr (down). Massive stars, $M > 40 M\odot$, are illustrated with larger symbols. Cluster members are color coded according to their velocities along the Galactic plane, $V_{x}$.}
\label{vel_xy}

\end{figure*}

\begin{figure*}

\centering $
\begin{array}{cc}
\includegraphics[trim=10mm 5mm 5mm 5mm, clip,width=95mm]{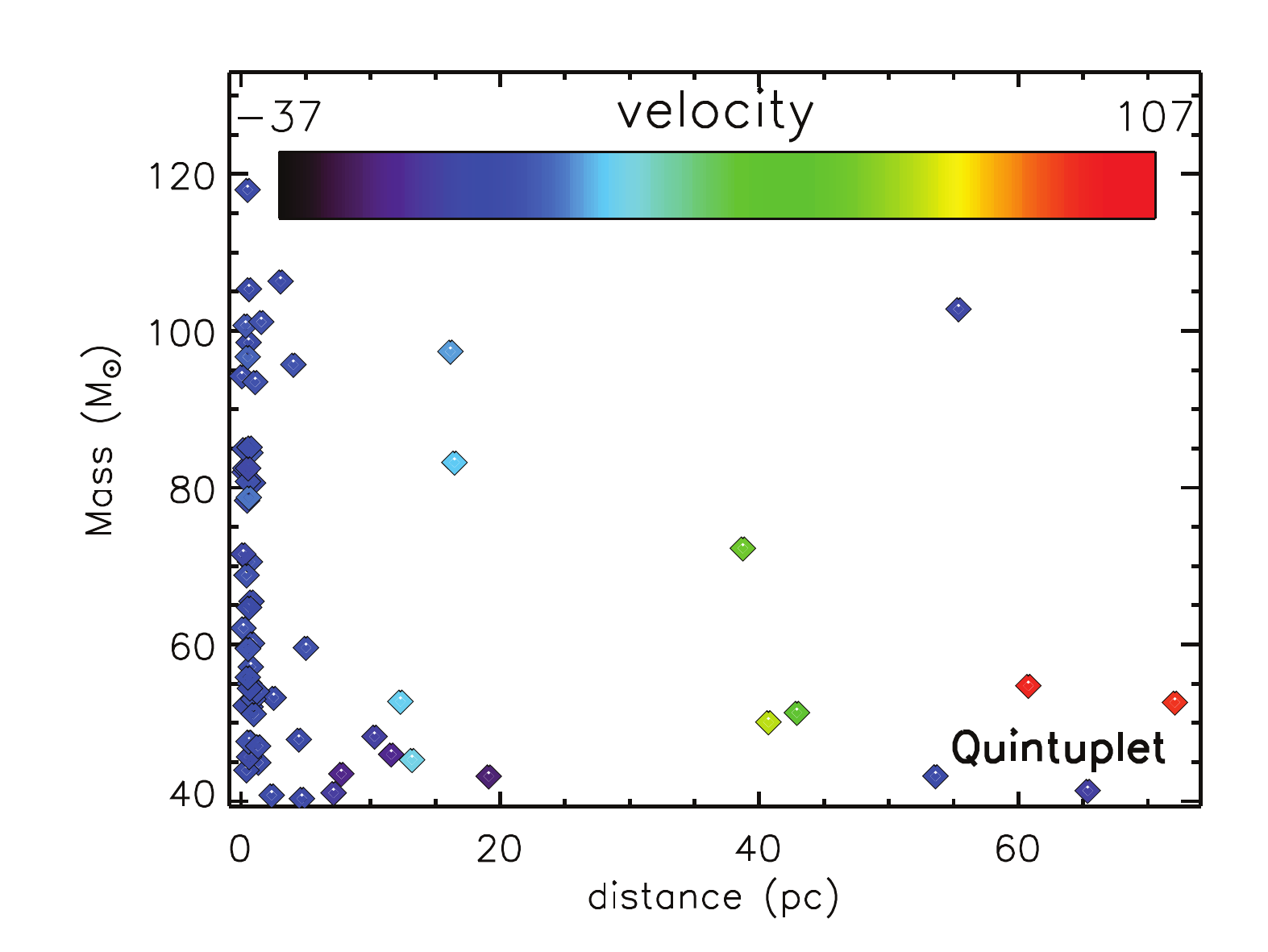} &
\includegraphics[trim=10mm 5mm 5mm 8mm, clip,width=95mm]{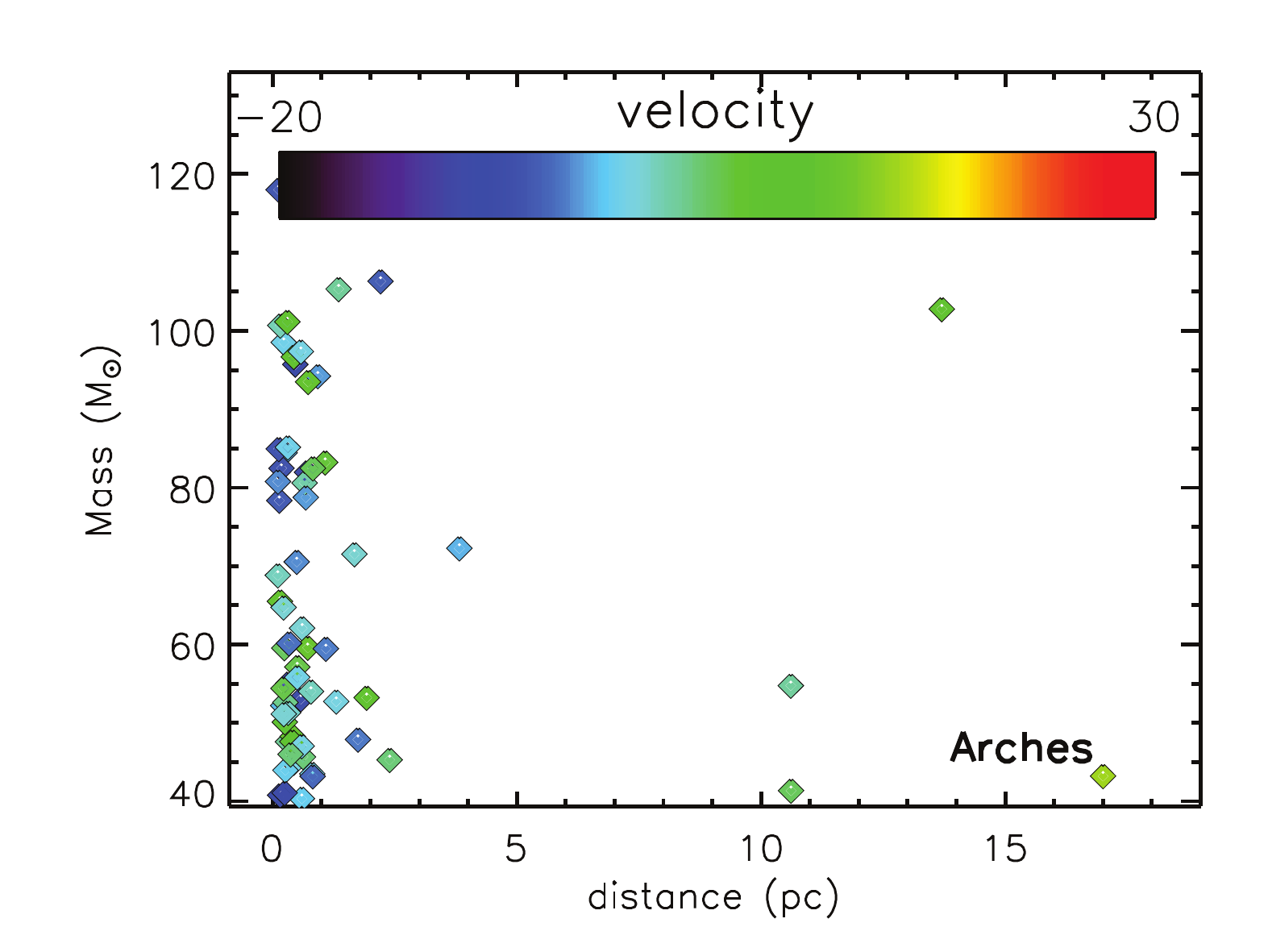} \\

\end{array}$
\caption{Initial masses are plotted over the distance from the cluster center for the massive stars, $M > 40 M\odot$. Stars are color coded  according to their velocities along the Galactic plane, $V_{x}$. Plotted populations are for one of the realizations of the model.}
\label{mass_d}
\end{figure*}

\begin{figure*}
\subfigure[]{\label{time_gcd}\includegraphics[trim=15mm 5mm 13mm 5mm, clip,width=94mm]{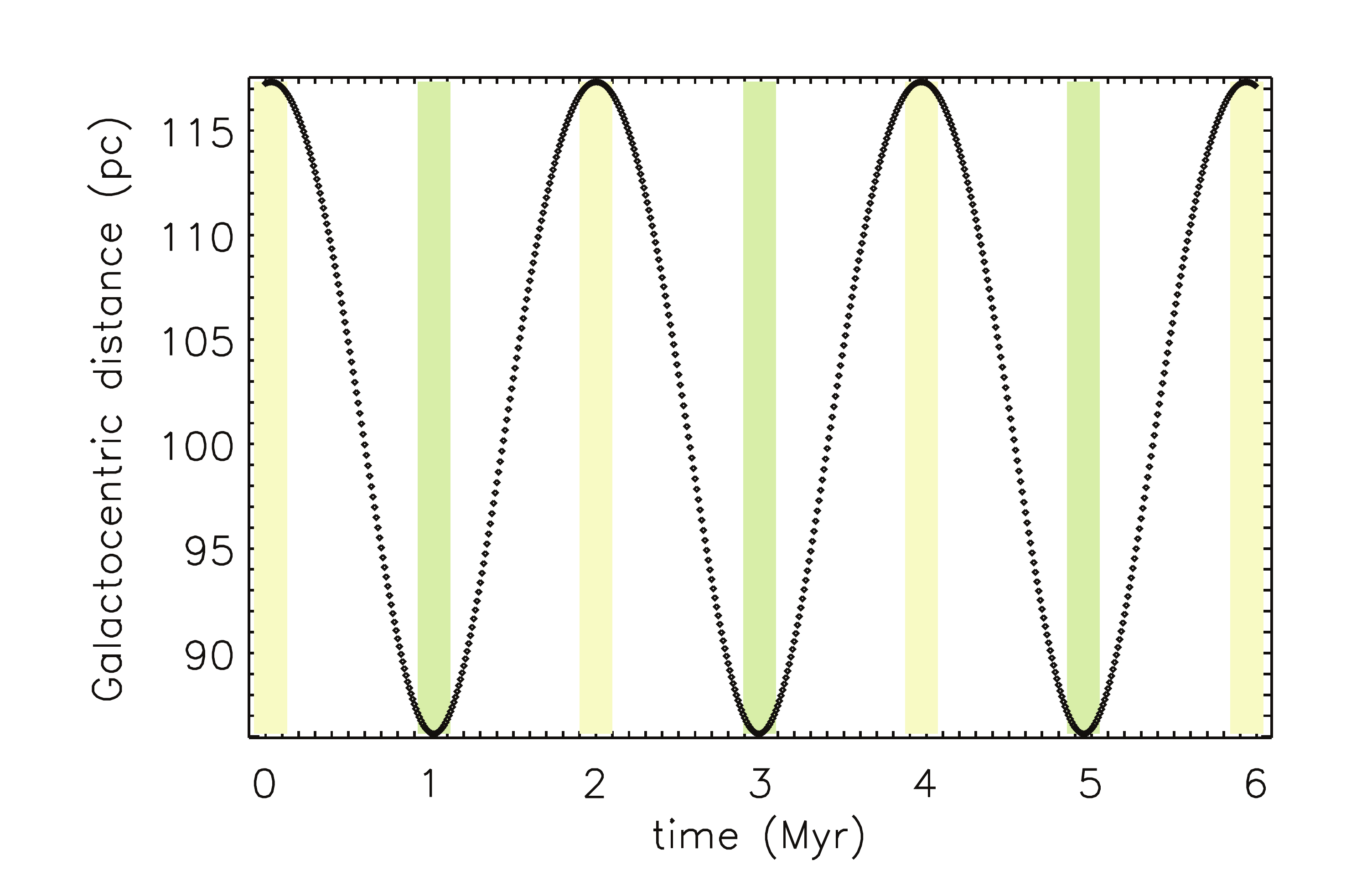}} 
\subfigure[]{\label{im_time_d}\includegraphics[trim=22mm 5mm 10mm 5mm, clip,width=94mm]{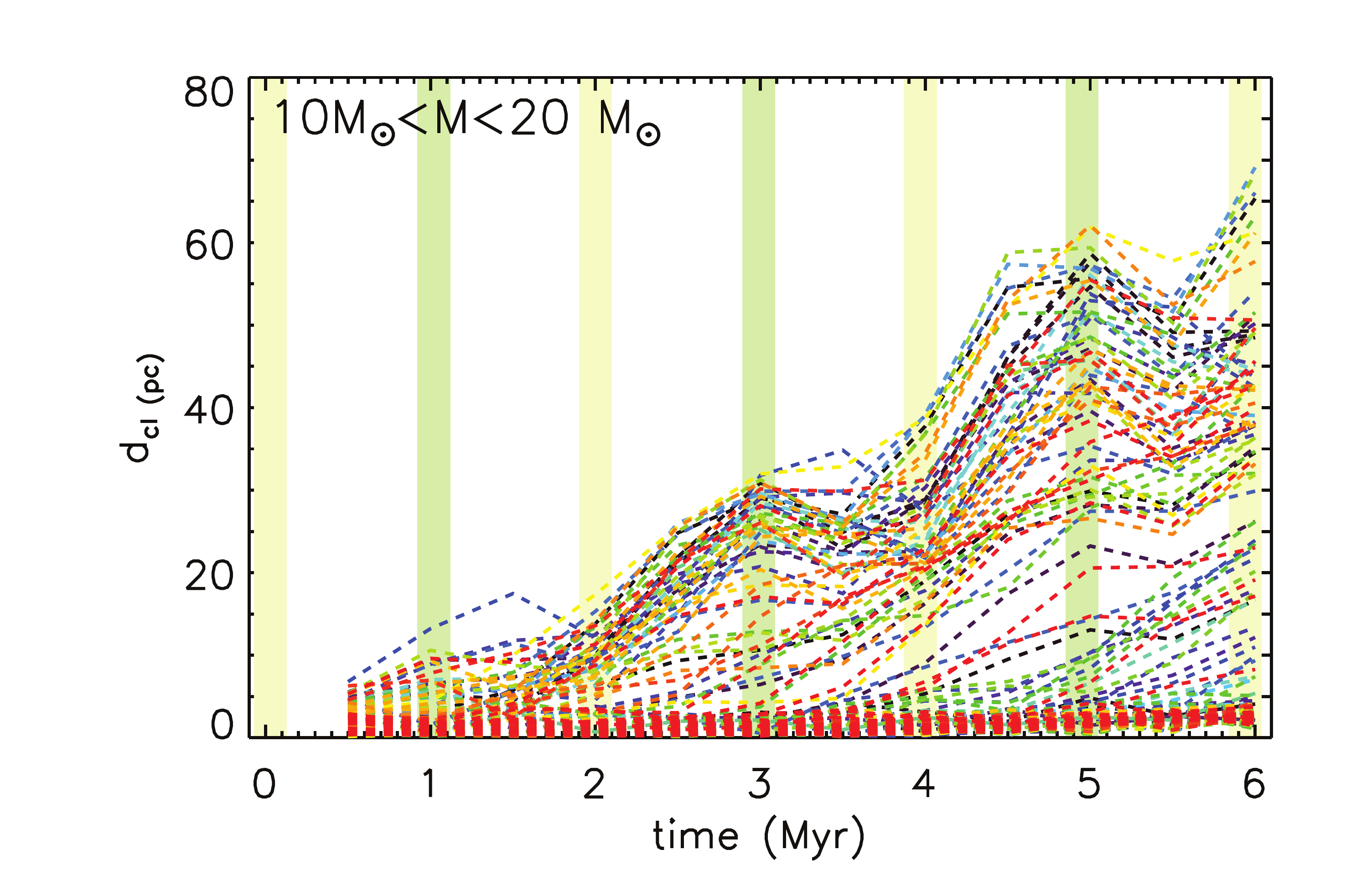}}
\subfigure[]{\label{hm_time_d}\includegraphics[trim=15mm 5mm 10mm 5mm, clip,width=94mm]{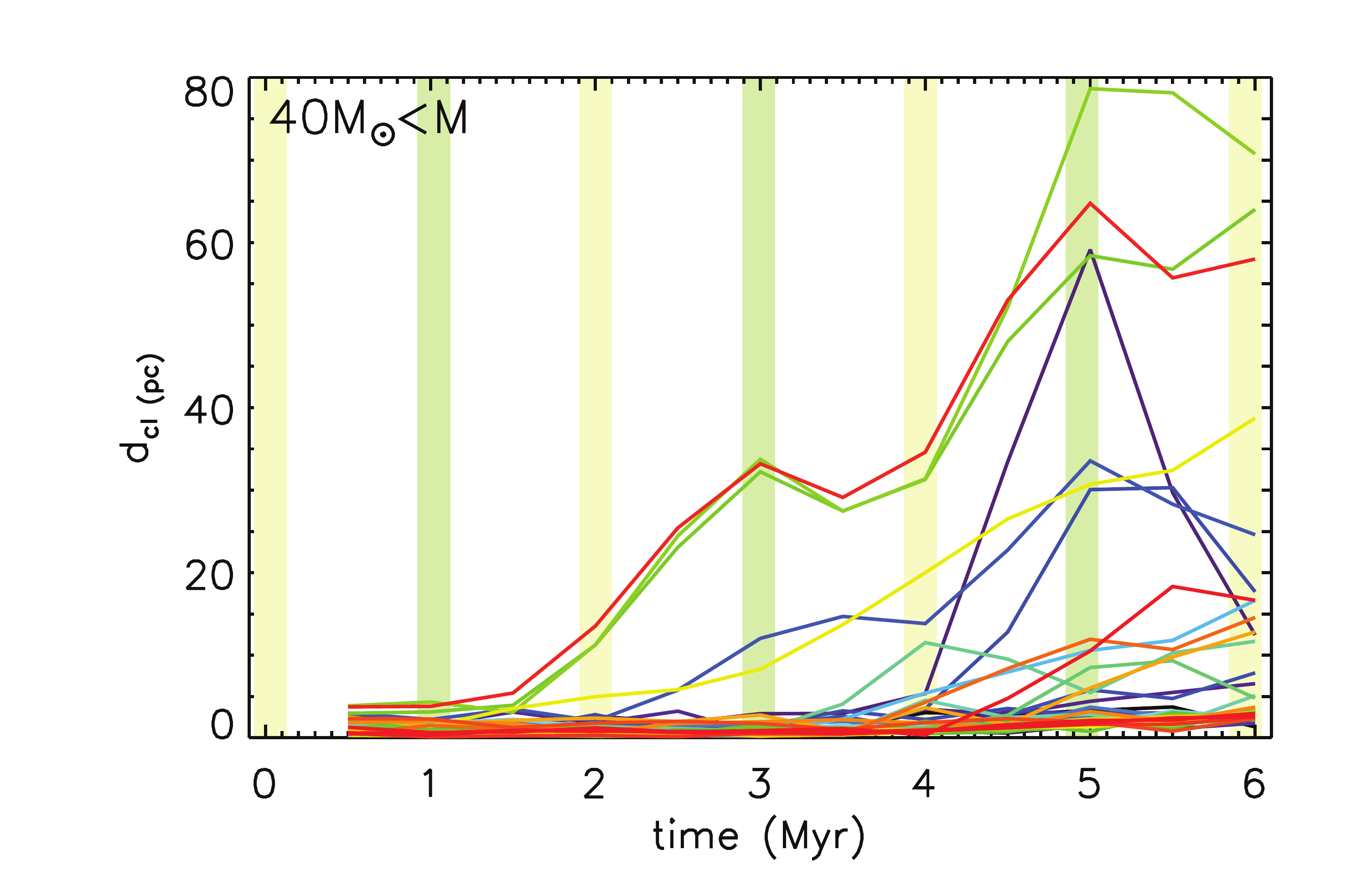}}
\subfigure[]{\label{density_core}\includegraphics[trim=0mm 5mm 10mm 5mm, clip,width=94mm]{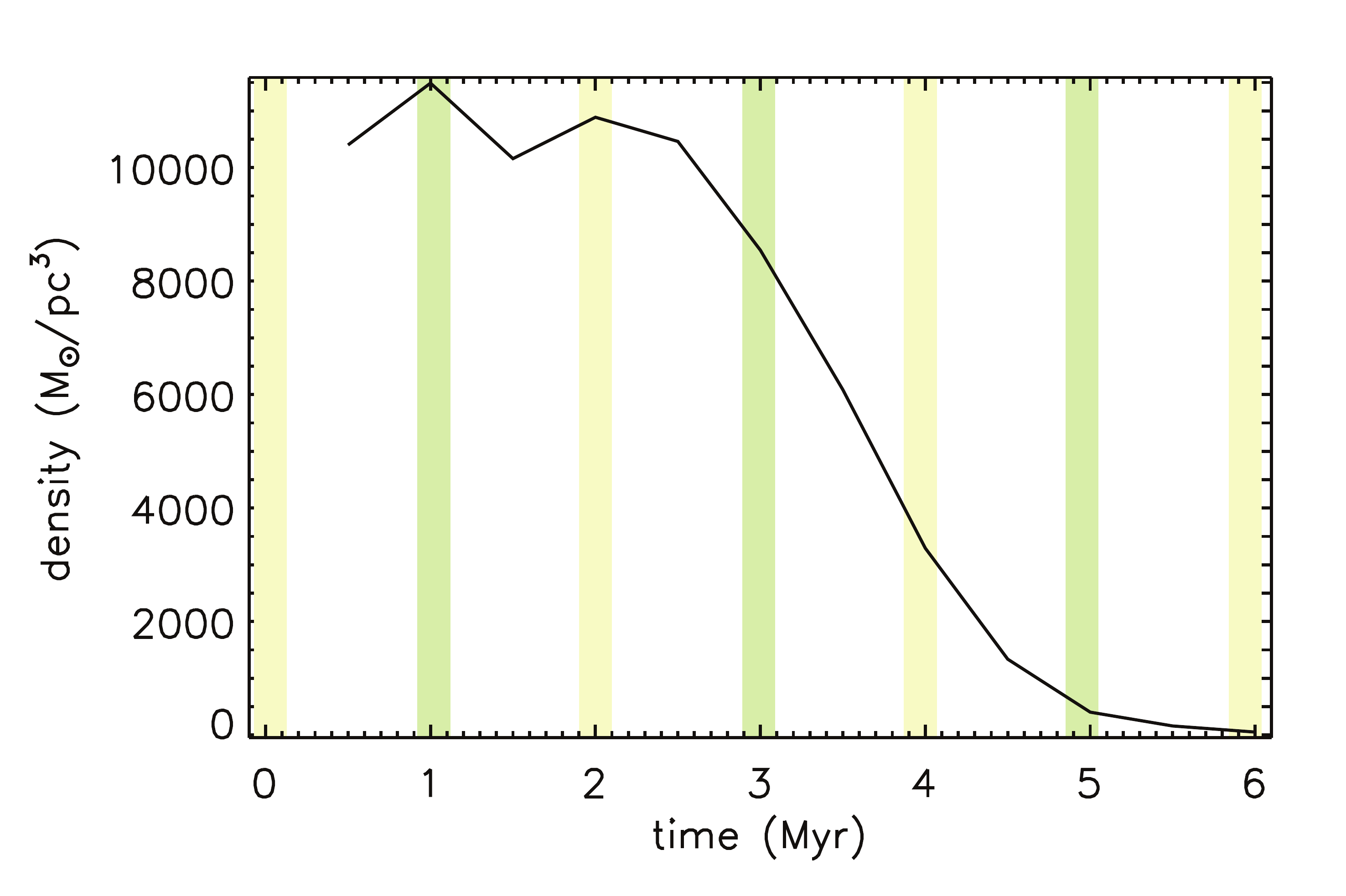}}
\caption{ (a): The projected galactocentric distance of the cluster on the plane of the sky (x,z) is plotted over time. As the cluster moves  along its open eccentric orbit, it encounters the variable tidal field caused by the GC. In figures (a) to (d) regions of maximum  and minimum Galactocentric distances are marked with yellow and green boxes respectively. (b, c): The projected distance of stars on the plane of the sky with reference to the cluster's center, $d_{cl}$, is plotted over time. Each line represents a trajectory of one star and colors are different to better distinguish lines.  Trajectories of the cluster members are illustrated for the two mass ranges of $10_{M\odot}< M < 20 \,M_{\odot}$  and  $M > 40 \,M_{\odot}$ in figures (b) and (c) respectively. (d): The surface mass density of the center of the cluster is plotted over time. The density is calculated for a cylinder of  $0.5\,\mathrm{pc}$ perpendicular to the plane of the sky.}
\label{time_d}

\end{figure*}

\begin{figure*}
 \centering
(a) \subfigure{\label{orbit_xy}\includegraphics[trim=10mm 5mm 12mm 12mm,clip,scale=0.25]{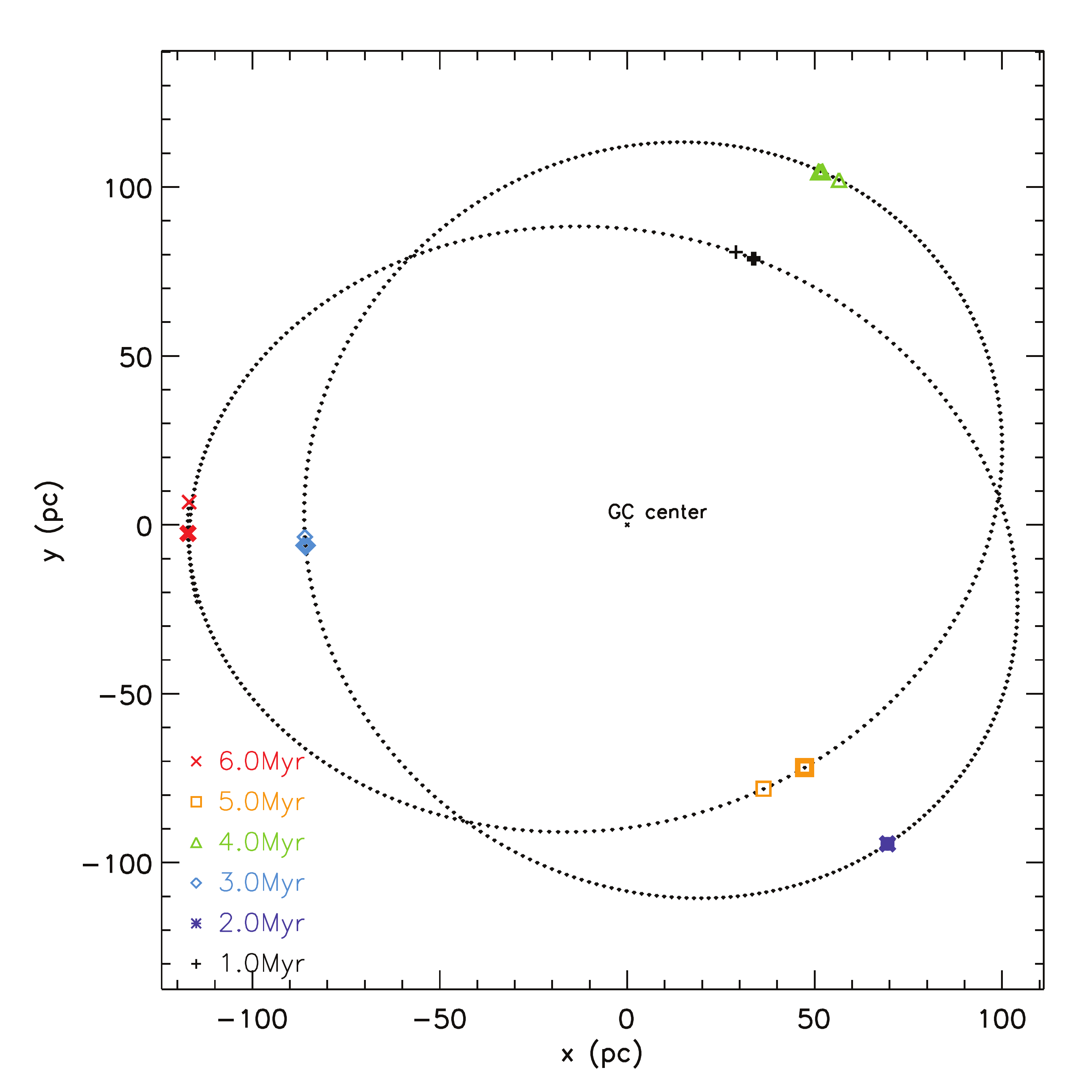}}\\
  \subfigure[]{\label{ss_xy}\includegraphics[trim=2mm 3mm 1mm 0mm,clip,scale=0.55]{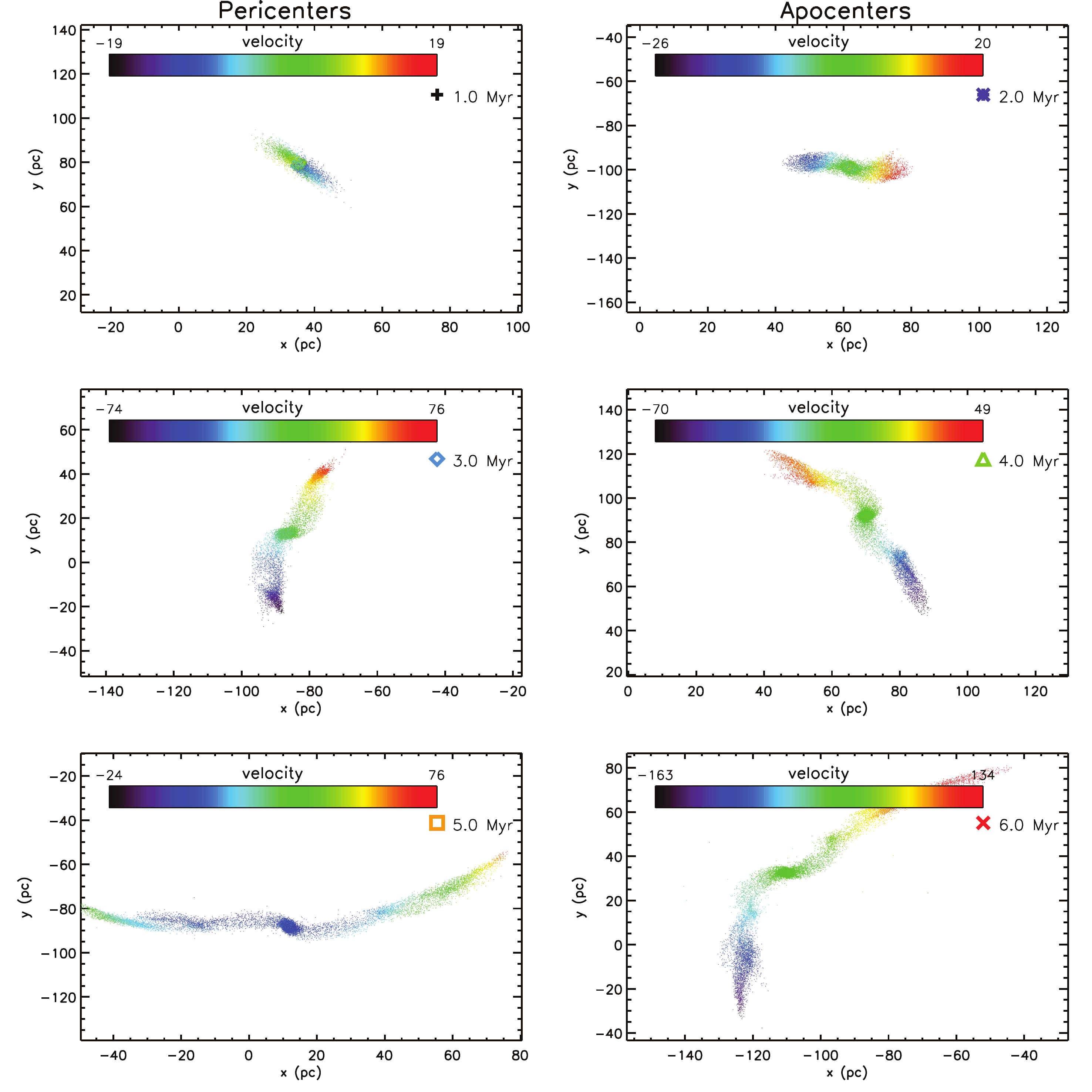}}
  \caption{ (a) The projected orbit of the cluster on the xy plane, which is close to the cluster's orbital plane, is illustrated. The symbols on the orbit mark the apsides of the 3D orbit. Each symbol/color corresponds to a particular age. The thick symbols mark the exact position of the apsides, while the thin symbols refer to the closest available snapshot of the simulated cluster considering the time resolution of the simulations. (b) The projected snapshots of the cluster on the xy plane are shown at apocenters and pericenters. Each snapshot corresponds to one marked apside on the projected orbit. The sources are colored based on the present velocity variation in the cluster and its tidal tails.}
\end{figure*}

Tidal arms form as a result of differential velocities produced by the many body dynamical interactions between stars in the cluster. Since the cluster moves along its orbit this velocity difference causes stars to either fall behind or  take over the cluster center. Figure  \ref{vel_xy} illustrates the velocity variation in the Arches (top panel) and the Quintuplet clusters (bottom panel). The velocity variation of the stars in the tidal arms reaches 50 $\mathrm{km\,s}^{-1}$ for the Arches cluster and 140 $\mathrm{km\,s}^{-1}$ for the Quintuplet cluster. The distribution of the massive sources, $M>40 \,M_{\odot}$, in Fig. \ref{mass_d} shows that despite the fact that most of the massive stars sink to the cluster core, in the furthest extent of the tidal tails there are sources with masses of up to $100 \,M_{\odot}$. For example in the  illustrated model in Fig. \ref{mass_d}, a $100 \,M_{\odot}$ star is found at the distance of $\sim 15\,\mathrm{pc}$ from the Arches and $\sim 60\,\mathrm{pc}$ from the Quintuplet center. Most of the massive sources which are outside the central tidal region have velocities similar to the tidally diffused population around them. The few massive sources that lie outside the tidal structure have velocities similar to the extreme velocities of stars in the tidal arms. To study the mechanism allowing the sources to drift
 out of the cluster in more detail, we analyze the trajectories of the sources in different mass ranges and compare them to the position of the cluster on its orbit around the GC. \cite{andrea2008} showed that the Arches cluster evolves along an open eccentric rosetta-like orbit (see Fig. \ref{orbit_xy} and  \ref{xz}(a)). Figure \ref{time_gcd} demonstrates the variation of Galactocentric distance of the cluster over time. As the cluster moves along its orbit, it encounters a spatially varying tidal field caused by the GC potential and its asymmetric orbit. The corresponding pericenters and the apocenters\footnote{As the cluster is not on an elliptical eccentric orbit we use the word pericenter (apocenter) only to refer to the point of closest (furthest) approach to the center of the potential on each orbit. The position of these points are different for each period.} on each period are indicated in figures \ref{time_d} and \ref{orbit_xy}. The projected orbit of the cluster in the xy plane, which is close to the orbital plane of the modeled cluster, is indicated in Figure \ref{orbit_xy}; the projected snapshots of the cluster at apsides are  illustrated in Fig. \ref{ss_xy} which allow us to follow the spatial spread and the velocity variation of the cluster along its orbit without being biased by the projection of the cluster on the plane of the sky, i.e., the xz plane.

  Following the pericenters and the apocenters on the orbit, we can trace the trajectory of the cluster members with respect to the cluster center (see Fig. \ref{time_d}). Both groups of massive stars, \mbox{$M > 40 M\odot$}, and stars with intermediate masses,  \mbox{$10 M_{\odot}< M < 20 \,M_{\odot}$}, follow a similar pattern (Figures \ref{im_time_d} and \ref{hm_time_d}). This similarity together with our finding that velocities of the simulated massive sources are in the range of the velocity variation along the tidal arms, are evidences that high-mass and intermediate-mass stars outside the tidal radius of the cluster gained energy through similar physical processes. Some of the stars gain energy in the center of the cluster probably through two-body relaxation. Afterward, as these sources recede from the cluster center, they undergo several periods of slightly falling  toward the cluster followed by pulling away from the cluster center. During the time that the cluster moves away from the apocenter toward pericenter its velocity increases, and consequently, individual orbits of stars on the tidal structures diverge spatially. This effect is indicated in Fig. \ref{time_d} as the sources recede from the cluster center after the apocenter passage. The projected snapshots of the cluster on the xy plane that is close to the orbital plane of the cluster illustrate the expansion of the cluster, including its tidal arms, after passing the apocenter and accelerating toward pericenter (see Fig. \ref{ss_xy}; compare the snapshots of the cluster on each pericenter to its snapshot at its previous apocenter). Comparably, when the cluster approaches apocenter and decelerates, stars on the tidal tails contract  towards the cluster center (see Figure \ref{time_d}) and the cluster appears less expanded (see Fig. \ref{orbit_xy}). A similar behavior is discussed in studies that investigate halo globular clusters and their interaction with the Galactic field through several Gyrs of evolution (e.g. \cite{2010MNRAS.401..105K}). We have calculated the surface mass density inside a cylinder of  $0.5\,\mathrm{pc}$ at the center of the cluster and perpendicular to the plane of the sky, as it can be investigated by observations. The surface mass density of the  central region of the cluster declines as the cluster expands on its orbit (see Fig. \ref{density_core}). 

 Although snapshots of the cluster along the xy plane, close to the orbital plane of the cluster, are instructive to follow the dynamical evolution of the cluster, the observed cluster will be affected by the projection of the cluster on the plane of the sky. Figure \ref{xz} in  Appendix A illustrates the projected orbit of the cluster on the plane of the sky together with snapshots of the cluster while crossing the apsides.
 
In summary, the models suggest that high-mass stars drift out of the clusters with the same physical mechanisms as intermediate-mass stars that form the tidal tails. Hence, their velocity is similar to the tail stars, and their location is consistent with the spatial distribution of the ``isolated'' WR stars close to the simulated location of the tidal tails. Figures \ref{im_time_d} and \ref{hm_time_d}, show that the drift pattern of high-mass stars over time is very similar to the drift pattern of intermediate mass stars. They follow the same apo/pericenter variation, and are not rapid ejectors in these models.  The clusters expand rapidly after 3 Myr, so that tidal tails stretch out over 120 pc across the central  molecular zone within 5 Myr (see Fig. \ref{xz} in Appendix A and Fig. \ref{ss_xy}), which is close to its entire diameter ($\sim$ 120 pc). According to these models, it is therefore not surprising anymore to find isolated high-mass stars at large distances from the location of the clusters today    as they drift along the tidal arms, they are expected to occupy the entire area where the tidal structures stretch out.

\section{Conclusions}\label{sec:conc}

In this study we present N-body simulations of the Arches cluster to create combined models of the Arches and Quintuplet clusters. The population of ejected and drifted sources from the two clusters is compared to the  HST/NICMOS  Paschen-$\alpha$ (Pa$\alpha$) survey of the Galactic center which detects the distribution of young massive stars in the GC region \citep{Mauerhan2010_main}.  Our study can be summarized as follows:

\begin{enumerate}
\item We construct different combined models of massive sources outside the Arches and Quintuplet clusters assuming different ages for the Quintuplet cluster and distinct values for the initial mass of a WR progenitor (see Table \ref{h_d}). We compare these models to the observed population of massive stars presented by \cite{Mauerhan2010_main}  employing a method which calculates the histogram difference between spatial distributions of stars. Among all the constructed models, the model which assumes an age of 5 Myr for the Quintuplet and an initial mass of $40 M_{\odot}$ for a WR progenitor is the most similar to the spatial distribution of the observed isolated high-mass stars. 

\item The strong tidal field of the GC potential results in extended tidal arms for both the Arches and Quintuplet clusters at their current age. In the best-matching model, tidal arms of the Arches cluster stretch out to 20 pc in each direction on the plane of the sky and along the Galactic plane, while the tidal arms of the Quintuplet cluster  extend out to 65 pc.  The observations of massive stars in the GC region \citep{Mauerhan2010_main} reveal two strips along the Galactic plane with a prominent gap  along the direction of the galactic poles. The projected positions of the tidal structures of the two clusters closely reproduce this observed distribution (see Fig. \ref{dens_map}).

\item The  observed massive sources outside the three clusters, including the young Nuclear cluster in the Pa$\alpha$ survey area is compared to the models \nnn{ using two different methods. First, comparing histograms of the spatial distributions of observed and simulated stars shows that the best-matching model reproduces $80\%$ of the observed population out to $21\,\mathrm{pc}$ and  $67\%$ of the observed population out to $80 \,\mathrm{pc}$ distance with reference to the center of the Arches cluster. Second, we create a density map of observed isolated massive stars using Voronoi diagrams. The constructed density map allows us to probe the probability of observing one or more stars in each Voronoi cell assuming our best-matching model.  For  62\% of the observed isolated massive stars, at least one of the ten random realizations of our model predicts  a star that can explain the observed star. This number increases to 72\% when we only consider the Voronoi cells within the central 20 pc from the center of the Arches cluster.} The sources that cannot be explained as originating from the Arches and Quintuplet are located at large distances
 from the tidal tails. On the other hand, the best-matching model predicts $20\%$ more massive stars outside the clusters, when we perform the comparison between the best-matching model and the list of observed WR stars only, i.e.,  excluding the less complete sample of OB sources (see Fig \ref{histo}).

\item The best-matching model predicts 26 massive stars outside the clusters, compared to the 35 observed massive stars outside the three clusters in the Pa$\alpha$ survey region. According to our best-matching models the majority of the simulated massive sources are located close to the tidal structure of the clusters, while the list of 35 observed massive stars includes 27 sources close to the tidal arms and 8 sources which are not close to the tidal structure of the Arches and the Quintuplet cluster. Histograms of the spatial distribution of the observed sources display a major characteristic peak at $\sim 15 \, \mathrm{pc}$ from the center of the Arches cluster. This peak is well reproduced in the best-matching model (see Fig. \ref{histo}). However, two minor peaks at the distances of $\sim 40\, \mathrm{pc}$ and  $\sim 60\, \mathrm{pc}$  from the Arches center are absent in the model. Possible origins of these sources are supernova kicks or dynamical ejections involving tight initial binary systems. In this case, these additional WR stars could also have emerged from the Arches and Quintuplet clusters. Currently we can also not exclude the possibility that these high-mass stars might have formed in smaller clusters or in isolation. Finally, a deviation in particular in the orbit of the Quintuplet cluster could also give rise to the remaining differences between the observed and the simulated spatial distribution of high-mass stars in the GC.

\item According to our models the projected tidal arms of the Quintuplet cluster at the age of 5 Myr extends out to $60\,\mathrm{pc}$ and reaches to the Sagittarius B2 region. This implies that the evolved massive stars observed in projection toward this region might originate from the tidally drifted sources from the Quintuplet cluster.

\item The tidal structure of both clusters form as a result of velocity variation in the cluster. The velocity variation along the tidal arms of the Arches cluster reaches $50\,\mathrm{km\,s}^{-1}$. This value is as high as 140 $\mathrm{km\,s}^{-1}$ for the Quintuplet cluster, that will be detectable in a proper motion diagram of high precision astrometric studies of the Quintuplet cluster.

\item The trajectories of the sources in different mass ranges in our models show that the tidal drifting of the cluster stars by the GC potential is an effective process which causes the stars to recede out to $70\,\mathrm{pc}$ from the center of the Arches cluster. Both massive, $M>40 \,M_{\odot}$, and intermediate mass stars, $10 M_{\odot}< M < 20 \,M_{\odot}$, follow a similar pattern; they gain energy in the center of the cluster which causes these sources to exceed the cluster's escape velocity. This suggests that the massive and intermediate-mass stars are evolved from the clusters by the same dynamical processes into tidal tails. The extended radial coverage of the high-mass stars inside the tidal tails implies that up to $80\%$ of the isolated observed WR population can be explained by cluster stars.

\end{enumerate}

\begin{acknowledgements}
      M.H. and A.S. acknowledge funding from the German science foundation (DFG) Emmy Noether program under grant STO 496-3/1, and thank the Argelander-Institute for Astronomy at the University of Bonn for being such a generous host. We wish to thank Dr. A. Liermann and Dr. C. Olczak for helpful discussions. We also thank Dr. C. Olczak for careful and substantial comments on the manuscript. \nnn{We thank  Dr. D. Applegate for valuable discussions on statistics. We are grateful to our anonymous referee for comments and suggestions that helped improve the paper.}
\end{acknowledgements}

\begin{appendix}
\section{The projected view on the plane of the sky}

\begin{figure*}[h]
 \centering
 \subfigure[]{\label{orbit_xz} \includegraphics[scale=0.6]{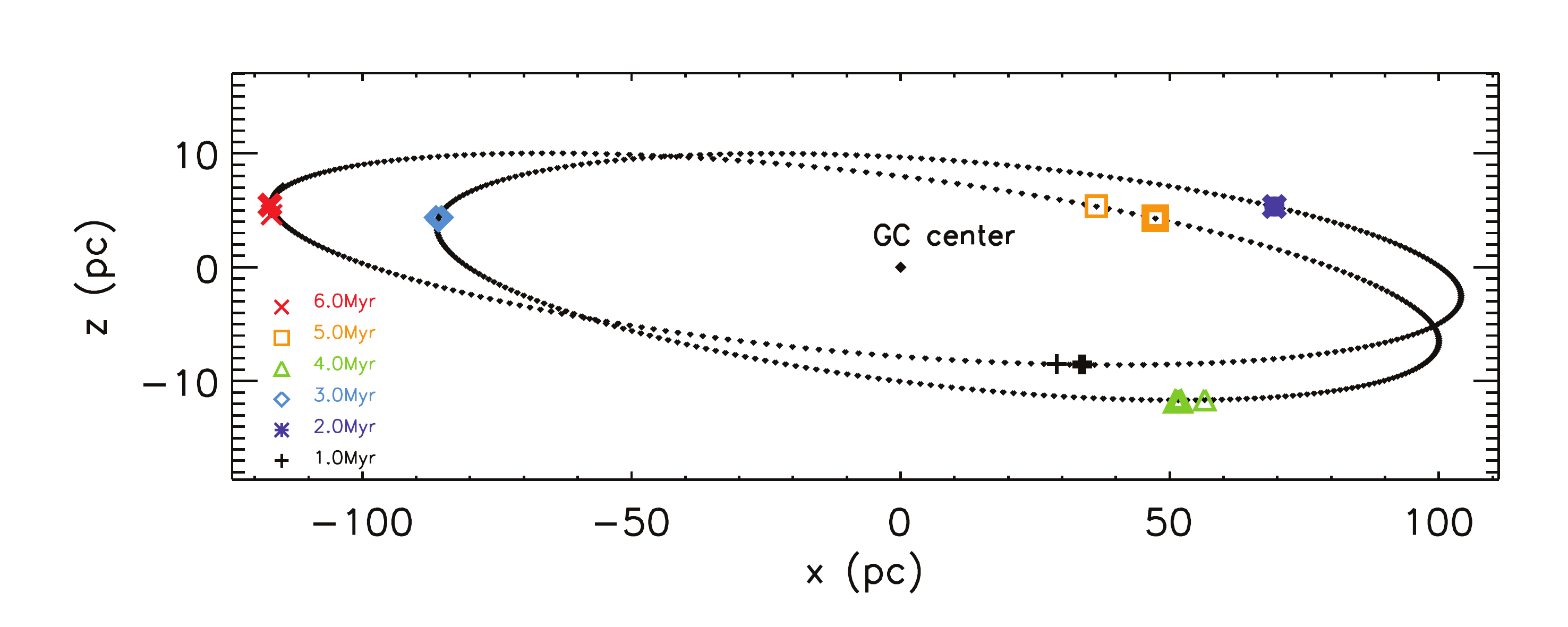}}
 \subfigure[]{\label{ss_xz} \includegraphics[trim=7mm 3mm 3mm 1mm, clip,scale=0.65]{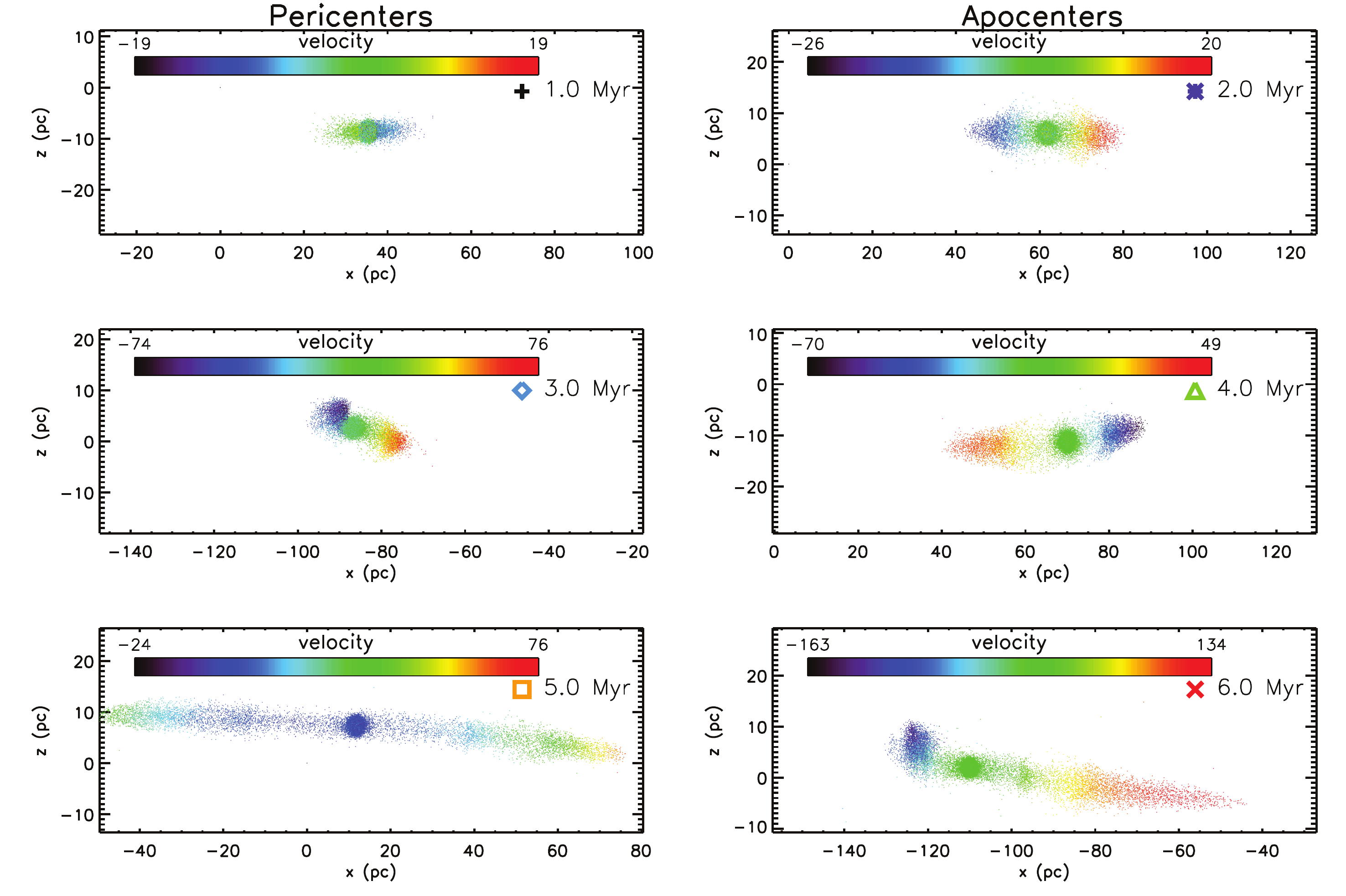}}
 \caption{(a) The projected orbit of the cluster on the plane of the sky, xz plane, is illustrated. The symbols on the orbit mark the apsides of the 3D orbit. Each symbol/color corresponds to a particular age. The thick symbols display the exact position of the apsides, while the thin symbols refer to the closest available snapshot of the simulated cluster considering the time resolution of the simulations. (b) The projected snapshots of the cluster on the plane of the sky are shown at apocenters and pericenters. Each snapshot corresponds to one marked apside on the projected orbit. The sources are colored based on the present velocity variation in the cluster and its tidal tails.} \label{xz}
\end{figure*}

In Fig. \ref{xz} we illustrate the projected orbit of the modeled cluster around the GC  on the plane of the sky, the xz plane. As the cluster moves along the 3D orbit it passes through apsides. These apsides are marked on the orbit together with the projected snapshot of the cluster at the corresponding apside point. Comparing Fig. \ref{xz}(b) with Fig. \ref{ss_xy} shows the strong effect of the  projection. For example the snapshot of the cluster at the age of $3\, \mathrm{Myr}$ appears squished on the plane of the sky (see Fig. \ref{xz}(b))  while comparing this snapshot to its projected snapshot on the xy plane (Fig. \ref{ss_xz}) reveals that the cluster is physically expanded at this location.\nnn{ The same effect is observed for some of our analyzed models.  Fig. \ref{orbit_xy} shows that at the age of 4.5 Myr the cluster is located in the middle of its two apside points, hence, it is  physically expanded. However, it appears squished on the plane of the sky. The effect of projection causes the massive stars in the tidal arms of the cluster  to spread over a much smaller area on the plane of the sky. Accordingly the $H_{d}$ value that reflects the similarity between the model and the observed population of isolated massive stars decrease for the model with an age of 4.5 Myr for the Quintuplet cluster (see Table \ref{h_d})}.

\end{appendix}

\bibliography{aa_paper.bbl}
\bibliographystyle{absrev}

\end{document}